\documentclass{article}
\usepackage{graphicx}
\usepackage{subcaption}
\usepackage{physics}
\usepackage{amsthm}
\usepackage{amsmath}
\usepackage{amsfonts}
\usepackage[affil-it]{authblk}
\usepackage{pict2e}
\usepackage{graphicx}
\usepackage{multirow}
\usepackage{svg}

\usepackage{tikz}
\usepackage{tikz-3dplot}
\usepackage{quantikz}
\usepackage{subcaption}

\usepackage{url}

\usepackage[T2A,T1]{fontenc}
\usepackage[utf8]{inputenc}
\usepackage[russian,english]{babel}
\usepackage{tempora}

\newcommand{\drawupstairs}[1]{
  \begin{tikzpicture}[x=0.5em, y=0.5em]
    \draw (0,0) \foreach \x in {0,...,#1} { -- (\x,\x+1) -- (\x+1,\x+1) };
  \end{tikzpicture}
}

\newcommand{\drawdownstairs}[1]{
  \begin{tikzpicture}[x=0.5em, y=0.5em]
    \draw (0, #1+1) \foreach \x in {0,...,#1} { -- (\x+1, #1+1-\x) -- (\x+1, #1-\x) };
  \end{tikzpicture}
}
\newtheorem{definition}{Definition}
\newtheorem{property}{Property}

\title{Stesso: A reconfigurable decomposition of $n$-bit Toffoli gates using symmetrical logical structures and adjustable support qubits}

\author[1,2]{Shanyan Chen}
\author[2]{Ali Al-Bayaty}
\author[2]{Xiaoyu Song}
\author[2]{Marek Perkowski}

\affil[1]{College of Information Engineering, Capital Normal University, China}
\affil[2]{Department of Electrical and Computer Engineering, Portland State University, USA}

\date{}

\begin{document}

\maketitle

\begin{abstract}
%introduction, problem, solution statement
    An $(n+1)$-bit Toffoli gate is mainly utilized to construct other quantum gates and operators, such as Fredkin gates, arithmetical adders, and logical comparators, where $n \geq 2$. Several researchers introduced different methods to decompose $(n+1)$-bit Toffoli gates in a quantum circuit into a set of standard 3-bit Toffoli gates or a set of elementary quantum gates, such as single-qubit and two-qubit gates.
    However, these methods are not effectively reconfigurable for linearly connected symmetrical structures (layouts) of contemporary quantum computers, usually utilizing more ancilla qubits. This paper introduces a new structural design method to effectively decompose $(n+1)$-bit Toffoli gates by utilizing configurable ancilla qubits, which we named the ``support qubits". Collectively, we call our decomposition method for symmetrical structures using support qubits the ``step-decreasing structures shaped operators (Stesso)".
    The main advantage of Stesso is to configurable construct different decomposed operators of various polarities and intermediate sub-circuits, such as Positive Polarity-Stesso, Mixed Polarity-Stesso, and Generalized-Stesso. With Stesso, it has been experimentally proven that $(n+1)$-bit Toffoli gates always have lower quantum costs than using conventional composition methods.
\end{abstract}


\section{Introduction}

%\textbf{remove anything related to IBM}

In quantum computing, the speed-up of all quantum search algorithms over classical search algorithms is essentially based on the superposition and entanglement of qubits~\cite{bub2010quantum,childs2018toward}. Compared with the quantum superposition state achieved by a single-bit Hadamard (H) gate, quantum entanglement is realized by $n$-bit quantum gates, such as the Feynman (CNOT or CX) gate for $n = 2$, and the Toffoli gate for $n \geq 3$. 
Given the physical connectivity requirements between entangled qubits, there will also be corresponding physical connectivity requirements between the control qubits and the target qubits in the $n$-bit gates, where $n \geq 2$.

% Among them, quantum entanglement requires connections between multiple entangled qubits.
Theoretically, all physical qubits can have full connectivity in a quantum processor unit (QPU). However, the connectivity between neighboring physical qubits of a QPU is limited by the physical limitations of its actual quantum layout (architecture)~\cite{larose2019overview,tan2020optimal}.
For instance, for different heavy-hex layouts of IBM QPUs~\cite{chamberland2020topological,ibm2025computer}, the number of neighboring physical qubits for each logical qubit in a circuit is often no more than four.
%, where the target (controlled) qubit is in the middle among the three control qubits. 
Experimentally, a set of decomposition methods for an $(n+1)$-bit Toffoli gate $\text{C}^{n}\text{X}$ ($n$ control qubits, one target qubit) becomes an important approach for the design of oracles in quantum search algorithms, where an oracle is the quantum circuit that conceptually describes the classical problems to be solved.
Notice that these methods become expensive and inefficient approaches, especially when $n$ increases.
In general, the 3-bit Toffoli gate is the so-called ``standard'' Toffoli gate of two controls and one target.

The research about decomposing an $(n+1)$-bit Toffoli gate started within two decades after the concept of the standard Toffoli gate was invented by Tommaso Toffoli~\cite{toffoli1980reversible} and is still ongoing. On the one hand, Barenco et al.~\cite{barenco1995elementary} demonstrated the decomposition of an $(n+1)$-bit Toffoli gate simulated by (i) a network consisting of $4(n-2)$ standard Toffoli (CCNOT or CCX) gates, and (ii) $48n-116$ elementary gates of single-bit rotational Pauli-Y ($\text{RY}(\frac{\pi}{4})$) and double-qubit gates of controlled-V and controlled-$\text{V}^\dagger$ with one garbage qubit. Additionally, Maslov and Dueck~\cite{maslov2003improved} designed $(n+1)$-bit Toffoli gates using Peres and inverse Peres gates. 

On the other hand, these two decomposition methods utilized the standard Toffoli gates. However, the mathematical proof of these designs is still missing,  along with texts and inspection.
Unlike using standard Toffoli gates, there are several decomposition designs of the $(n+1)$-bit Toffoli gate consisting of elementary gates. For example, Orts et al.~\cite{orts2022studying} studied the cost of the $(n+1)$-bit Toffoli gate, mainly focusing on the count of the rotational Pauli-Z gates $\text{RZ}(\frac{\pi}{4})$ (T gates) and the number of utilized ancilla qubits. Recently, Claudon et al.~\cite{claudon2024polylogarithmic} introduced an approximating decomposition without ancilla and two exact decompositions, one with one borrowed ancilla qubit, while another with a reduced number of ancilla qubits.

The goal of this paper is to introduce a new structural design methodology of $(n+1)$-bit Toffoli gates using our three proposed construction advantages:
\begin{enumerate}
    \item Reconfigurability, by adding and removing different Boolean operations (such as negation) and constrained unitary gates (such as CX gates).
    \item Adjustable support qubits, a set of adjustable ancilla qubits whose initial states don't matter, which are realized by reusing a subset of the input qubits of an $(n+1)$-bit Toffoli gate.
    \item Layout-awareness, which is well-suited for mapping (routing) to specific linearly connected symmetrical structures~\cite{QiskitLattice,yu2024symmetry} of QPUs, such as square lattices~\cite{garvin2025data}.
    % triangular layouts~\cite{yang2025new}
\end{enumerate}

Based on these three construction advantages, we effectively decomposed an $(n+1)$-bit Toffoli gate into different symbolically shaped logical structures of a set of standard Toffoli gates and X gates, such as English-shape (like V-shape), Cyrillic-shape (like {\foreignlanguage{russian}{И}}-shape), and symbol-shape (like $\backslash$-shape).
% V-shaped uniform structure of a set of standard Toffoli gates. 
We called this structure the ``step-decreasing structures shaped operators (Stesso)",
 with positive polarities ``PP-Stesso", mixed polarities ``MP-Stesso", and generalized ``G-Stesso" of mixed SOP and ESOP logical structure.
 Notice that SOP is a fundamental structural form for classical design, while ESOP is a fundamental structural form for quantum design.
Experimentally, an $n$-bit binary comparator, which consists of $(n+1)$-bit Toffoli gates and our Stesso, was designed and evaluated. The Results and Discussions section proves that the Stesso is always more efficient than using $(n+1)$-bit Toffoli gates.

\section{Background and related work}

\subsection{Decomposition methods for $\text{C}^n\text{X}$ gates}

There are two general decomposition methods for $\text{C}^n\text{X}$ gates distinguished by using three-qubit gates (like the standard 3-bit Toffoli gates and Peres gates) or not, where $n \geq 3$.
On the one hand, Barenco et al.~\cite{barenco1995elementary} demonstrated the decomposition of an $(n+1)$-bit Toffoli gate simulated by (i) a network consisting of $4(n-2)$ standard Toffoli (CCNOT or CCX) gates, and (ii) $48n-116$ elementary gates of single-bit rotational Pauli-Y ($\text{RY}(\frac{\pi}{4})$) and double-qubit gates of controlled-V and controlled-$\text{V}^\dagger$ with one garbage qubit. Additionally, Maslov and Dueck~\cite{maslov2003improved} designed $(n+1)$-bit Toffoli gates using Peres and inverse Peres gates. 

On the other hand, these two decomposition methods utilized the standard Toffoli gates. However, the mathematical proof of these designs is still missing,  along with texts and inspection.
Unlike using standard Toffoli gates, there are several decomposition designs of the $(n+1)$-bit Toffoli gate consisting of elementary gates. For example, Orts et al.~\cite{orts2022studying} studied the cost of the $(n+1)$-bit Toffoli gate, mainly focusing on the count of the rotational Pauli-Z gates $\text{RZ}(\frac{\pi}{4})$ (T gates) and the number of utilized ancilla qubits. Recently, Claudon et al.~\cite{claudon2024polylogarithmic} introduced an approximating decomposition without ancilla and two exact decompositions, one with one borrowed ancilla qubit, while another with a reduced number of ancilla qubits.

All above-mentioned decomposition methods are only useful to generate an $(n+1)$-bit Toffoli gate with positive polarities of all control qubits. Different from these methods, our newly proposed structure design methodology is able to support mixed polarities of sup-product terms of control qubits and specific output forms of mixed OR and ESOP while maintaining its implementing complexity from $O(log(n))$ to $O(n)$.

\subsection{Binary comparator}

An $n$-bit binary comparator~\cite{mano2023digital}, which is also called a digital comparator and a binary magnitude comparator, compares two $n$-bit numbers $x_n \cdots x_1$ and $y_n \cdots y_1$, and produces three 1-bit outputs whether the first $n$-bit number is less than, equal to, or greater than the second $n$-bit number.
The design of quantum binary comparators is related to Boolean-based arithmetic and reversible logic. In classical computing, comparators are basic components of Arithmetic Logic Units (ALUs) and form the basis of conditional statements. However, translating this functionality to the quantum realization presents unique challenges, including managing fragile quantum states, optimizing resource usage (qubits and gates), and designing for scalability~\cite{tyagi2020high}.

Early research focused on emulating classical sequential logic in the quantum domain, such as the iterative quantum comparator proposed by Al-Rabadi~\cite{al2009closed}. While straightforward, his method requires a large number of ancillary qubits that scale with the input size, and it suffers from high circuit latency or quantum depth. To reduce quantum delay, researchers later developed more parallel, tree-based architectures~\cite{thapliyal2010design}. 
This methodology significantly reduced latency compared to serial designs, but generally at the expense of more ancillary qubits.
%Unlike these two methodologies, 
% Another different approach is QFT-based subtraction approach~\cite{yuan2023improved}. The QFT-based method is also effective for comparing a quantum state against a classical number, a common requirement in many algorithms.
% Recent advancements focus on developing more generalized and resource-efficient comparators.
For fault-tolerant quantum computing, where T-gates are computationally expensive, research has focused on optimizing comparator circuits to minimize the use of these T-gates, such as in a recent paper~\cite{donaire2024lowering} about the T-gate optimization.
However, their design methodology requires several number of ancilla qubits, as expressed in Table 8 in~\cite{donaire2024lowering}.

Unlike these methods, an $n$-bit binary comparator can be directly designed by using our Stesso without auxiliary ancilla qubits, i.e., only $2n$ inputs and three outputs involved.

% This involves efficient decomposition of multiple-controlled gates and reusing ancilla qubits.
% Recently, 

% uses the quantum Fourier transform (QFT) to perform arithmetic operations, which can be adapted for comparison.

% For example, Proposed an improved QFT-based quantum comparator that is highly resource-efficient, using only a single ancilla qubit.

% There are three methodologies for designing an $n$-bit quantum binary comparator, i.e., sequential (bit-by-bit) comparison [XXX], subtraction-based comparison [XXX], and Quantum Fourier Transform (QFT)-based comparison [XXX].

\section{Methods}

This section introduces our methodology for effectively decomposing $(n+1)$-bit Toffoli gates. Firstly, we give the reason why our newly introduced step-decreasing structure shaped operators (Stesso) are well-suited for symmetrical structures and why the Stesso consists of standard Toffoli gates instead of elementary gates by discussing suitable actual symmetrical layouts.
Secondly, we define the symbol-shaped structures, English-shaped structures, and Cyrillic-shaped structures as the basic logically shaped structures for Stesso.
Finally, we define the Stesso for an $(n+1)$-bit Toffoli gate based on these different basic shapes.
% Finally, to further enhance the composability of Stesso, we present local transformations and rules for composing more than one $(n+1)$-bit Toffoli gate, which can recursively create a final Stesso.

% Secondly, we define the Stesso for an $(n+1)$-bit Toffoli gate based on different shapes of the final composed standard Toffoli gates.
% Finally, to further enhance the composability of Stesso, we present local transformations and rules for composing more than one $(n+1)$-bit Toffoli gate, which can recursively create a final Stesso.

% To further reduce the cost of quantum gates in the combined circuit composed of multiple Stessos, 
% to enhance the understanding of the physical connectivity requirements and symmetrical structures, we give the reason why our newly proposed step-decreasing structure shaped operators (Stesso)
% the layout-awareness construstion 

% Firstly, this section discusses suitable triangular layouts and symmetrical lattices to enhance the understanding of the physical connectivity requirements for control qubits and the target qubit.
% Then, it further presents our methodology for effectively decomposing $(n+1)$-bit Toffoli gates using:
% \begin{enumerate}
%     \item Step-decreasing structures shaped operators (Stesso) for one $(n+1)$-bit Toffoli gate.
%     \item Local transformations and rules for more than one $(n+1)$-bit Toffoli gates and creating one final Stesso recursively.
% \end{enumerate}

% \textbf{Move it to the result and discussion}
\subsection{Symmetrical structures}

In quantum computing, a standard Toffoli gate requires the full physical connectivity among two control qubits and one target qubit. Obviously, for all possible combinations of quantum circuits consisting of standard Toffoli gates, the physical connectivity requirement is a set of all the full physical connectivity required of all involved standard Toffoli gates. Therefore, according to the set theory~\cite{hausdorff2021set}, the number of appearances of standard Toffoli gates with the same control qubits and target qubit in different stages of a quantum circuit only counts once.

\begin{figure}[htbp]
    \centering
    \begin{subfigure}[b]{0.3\textwidth}
        \includegraphics[width=\textwidth]{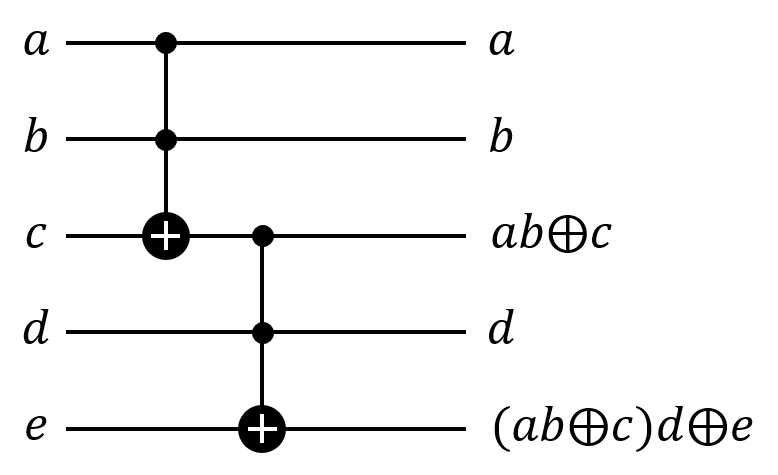}
        \caption{symbol-shape structure ($\backslash$-shape)}
        \label{fig:Toffoli_exp1}
    \end{subfigure}
    %\hfill
     \hspace{0.05\textwidth}
    \begin{subfigure}[b]{0.3\textwidth}
        \includegraphics[width=\textwidth]{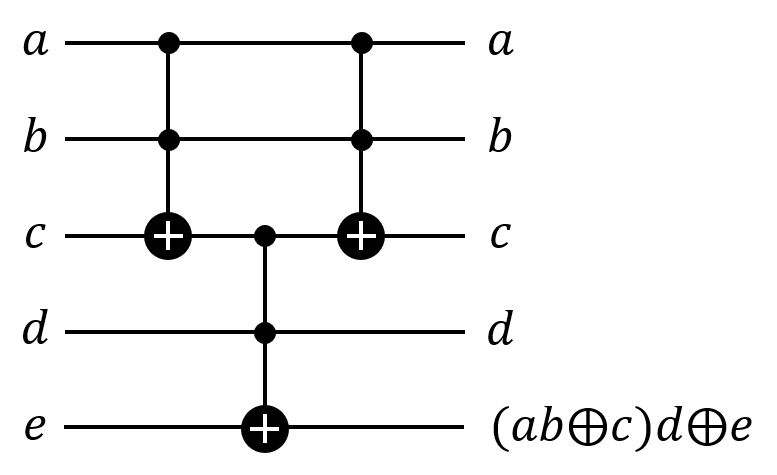}
        \caption{English-shape structure (V-shape)}
        \label{fig:Toffoli_exp2}
    \end{subfigure}
\hspace{0.05\textwidth}
    \begin{subfigure}[b]{0.2\textwidth}
        \includegraphics[width=0.8\textwidth]{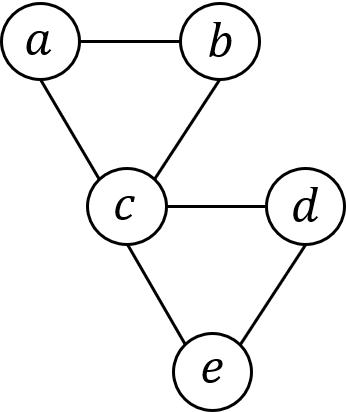}
        \caption{Connectivity requirements}
        \label{fig:Triangle_layout1}
    \end{subfigure}
    \caption{Examples of combined quantum circuits of standard Toffoli gates with the same physical connectivity requirement for qubits $a,b,c,d,e$, utilizing symbol-shape and English-shape structures of Stesso.}
    \label{fig:Toffoli_exp}
\end{figure}

For example, there are two distinct-shaped structures of Stesso, as illustrated in Figure~\ref{fig:Toffoli_exp}. These two shapes have the same physical connectivity requirements of all involved qubits, as shown in Figure~\ref{fig:Triangle_layout1}.
% In Figure~\ref{fig:Toffoli_exp}, the connectivity requirements of these two combined quantum circuit examples of standard Toffoli gates are the same, as shown in Figure~\ref{fig:Triangle_layout1}.
Notice that the connectivity requirements in Figure~\ref{fig:Triangle_layout1} are well-suited to a triangular layout.

% \begin{figure}[htbp]
%     \centering
%     \begin{subfigure}[b]{0.2\textwidth}
%         \includegraphics[width=\textwidth]{figs/Layout-Triangle-Map.png}
%         \caption{Connectivity requirements}
%         \label{fig:Triangle_layout1}
%     \end{subfigure}
%     %\hfill
%      \hspace{0.1\textwidth}
%     \begin{subfigure}[b]{0.25\textwidth}
%         \includegraphics[width=\textwidth]{figs/Layout-Triangle.png}
%         \caption{Triangular layout}
%         \label{fig:Triangle_layout2}
%     \end{subfigure}
%     \caption{Physical connectivity requirements of different shapes of standard Toffoli gates and the well-suited triangular layout.}
%     \label{fig:Triangle_layout}
% \end{figure}

When the quantum layout is not the same as the connectivity based on the triangle structure in Figure~\ref{fig:Triangle_layout1}, it is necessary to find an effective mapping (rerouting) method to map logical qubits to physical qubits to meet their physical connectivity requirements. This mapping method is limited by the connectivity number and location of the neighboring physical qubits of an actual QPU, i.e., different mapping methods for different actual quantum layouts.

\begin{figure}[htbp]
    \centering
    \begin{subfigure}[b]{0.15\textwidth}
        \includegraphics[width=\textwidth]{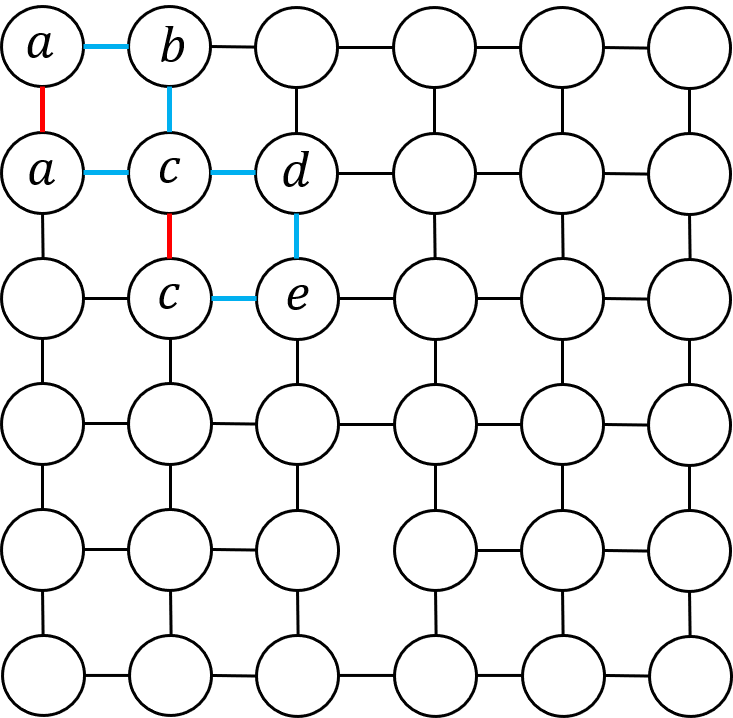}
        \caption{Rigetti}
        \label{fig:layout-rigetti-map}
    \end{subfigure}
    %\hfill
     \hspace{0.05\textwidth}
    \begin{subfigure}[b]{0.3\textwidth}
        \includegraphics[width=\textwidth]{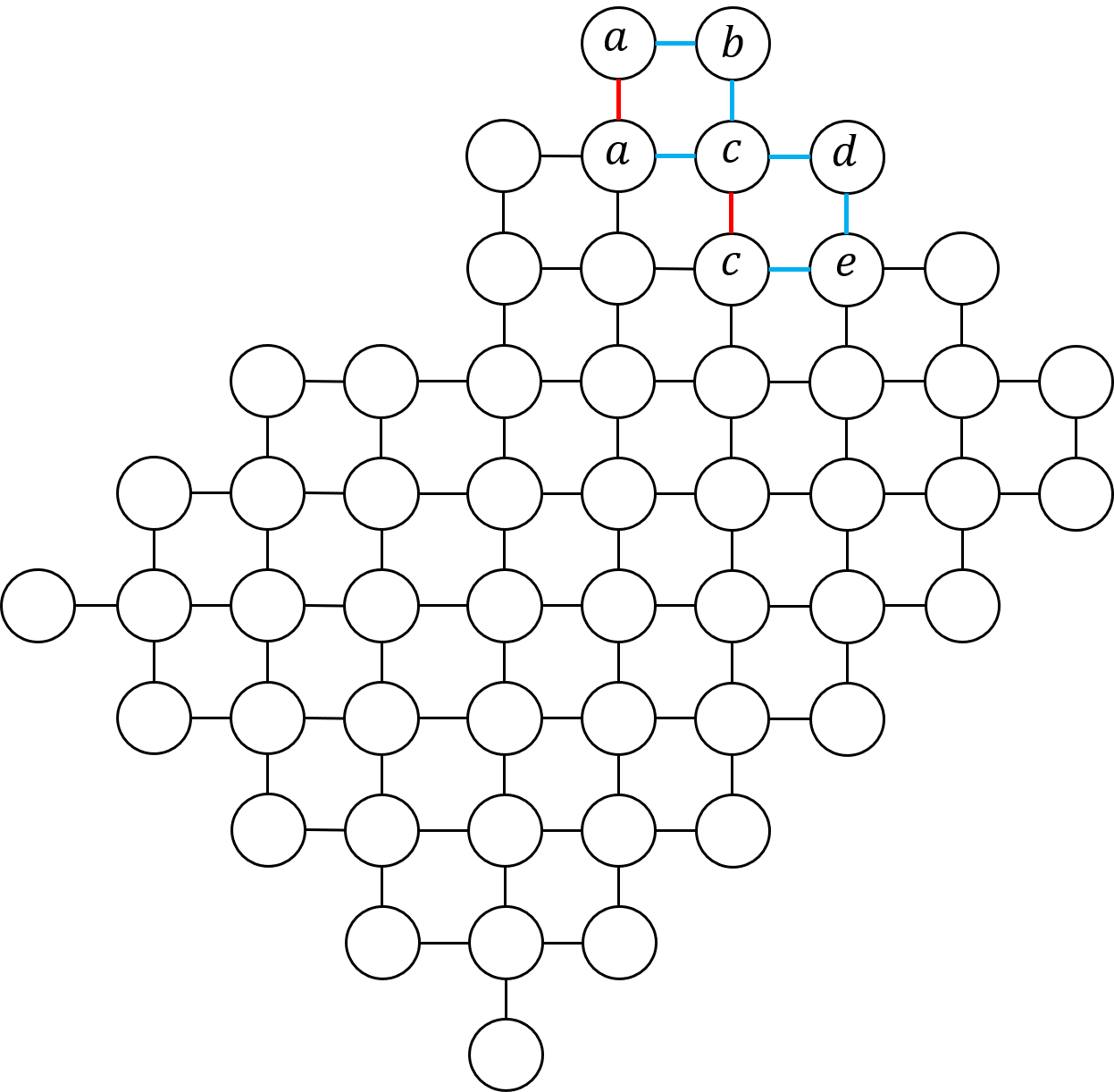}
        \caption{Google}
        \label{fig:layout-google-map}
    \end{subfigure}
    \hspace{0.05\textwidth}
    \begin{subfigure}[b]{0.4\textwidth}
        \includegraphics[width=\textwidth]{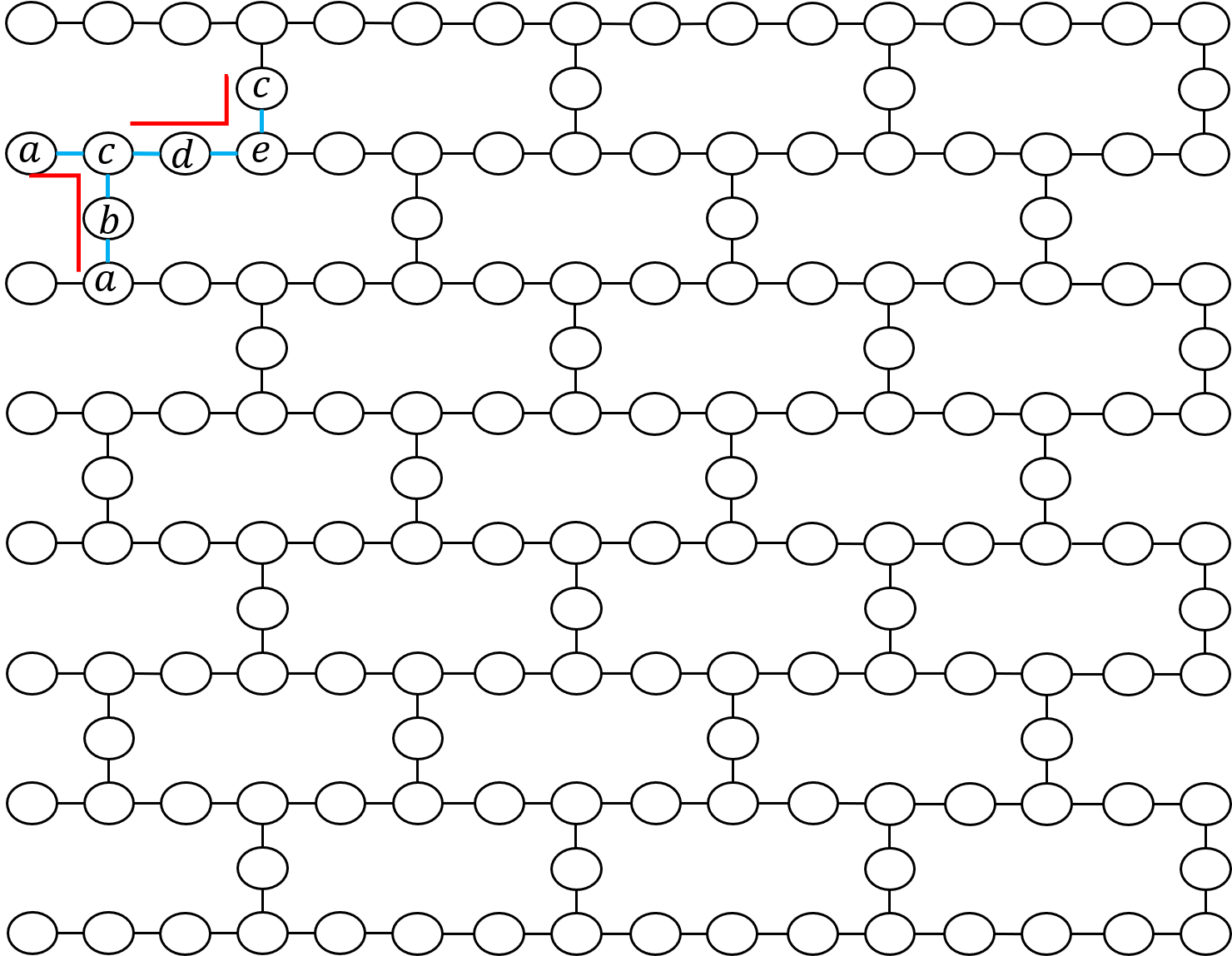}
        \caption{IBM}
        \label{fig:layout-IBM-map}
    \end{subfigure}
    \caption{Possible mapping results of qubits $a,b,c,d,e$ in the quantum structures of Rigetti, Google, and IBM QPUs meeting the connectivity requirements in Figure~\ref{fig:Triangle_layout1}. The red line (bus) indicates that SWAP gates are added based on the distance between two physical qubits. Blue line (bus) means the direct physical connection between two physical qubits. Circles mean the physical qubits of a QPU. The letters ($a,b,c,d,e$) are the logical qubits of standard Toffoli gates.}
    \label{fig:sym_structs_map}
\end{figure}

Notice that the current existing basic quantum structures used in actual QPUs generally have symmetry, such as square lattices of Rigetti~\cite{rigetti2025QCS,garvin2025data} in Figure~\ref{fig:layout-rigetti-map}, square-grid structures of Google~\cite{arute2019quantum,google2025quantum} in Figure~\ref{fig:layout-google-map}, heavy-hex lattices of IBM~\cite{ibm2025computer,chang2021mapping,hetenyi2024creating} in Figure~\ref{fig:layout-IBM-map}. In this paper, the symmetry of a structure is the mathematical ``translation symmetry", which means a symmetrical structure remains the same edge-vertex connectivity after being ``shifted or slid" in a specific direction by a fixed distance.

% \begin{figure}[htbp]
%     \centering
%     \begin{subfigure}[b]{0.15\textwidth}
%         \includegraphics[width=\textwidth]{figs/Layout-Square-Rect.png}
%         \caption{Rigetti}
%         \label{fig:layout-rigetti}
%     \end{subfigure}
%     %\hfill
%      \hspace{0.05\textwidth}
%     \begin{subfigure}[b]{0.3\textwidth}
%         \includegraphics[width=\textwidth]{figs/Layout-Sycamore.png}
%         \caption{Google}
%         \label{fig:layout-google}
%     \end{subfigure}
%     \hspace{0.05\textwidth}
%     \begin{subfigure}[b]{0.4\textwidth}
%         \includegraphics[width=\textwidth]{figs/Layout-Heavy-hex.png}
%         \caption{IBM}
%         \label{fig:layout-IBM}
%     \end{subfigure}
%     \caption{Symmetrical quantum structures (layouts) of various QPUs: (a) mixed rectangular and square lattices in Rigetti's Cepheus-1-36Q QPU~\cite{rigetti2025QCS,garvin2025data}, (b) square-grid structures in Google's Sycamore QPU~\cite{arute2019quantum,google2025quantum}, and (c) heavy-hex lattices in IBM's Pittsburgh QPU~\cite{ibm2025computer,chang2021mapping,hetenyi2024creating}.}
%     \label{fig:sym_structs}
% \end{figure}

% Mathematically, symmetry for groups and graphs refers to automorphisms, and symmetry for lattices is translation symmetry or rotation symmetry. Simply from the shape, generally, having the space at the center symmetry of the structure can be called a symmetric structure.

Obviously, the triangular structure of standard Toffoli gates in Figure~\ref{fig:Triangle_layout1}, which meets the full physical connectivity requirements, is symmetrical for both translation and rotation. 
The rotation symmetry means the structure remains identical (same) after being rotated a specific angle at the center of this structure.
Therefore, the Stesso introduced in the next section also has the same symmetrical property. For this reason, our Stesso is well-suited for mapping to the quantum layouts based on symmetrical structures.
Specifically, for the actual symmetrical quantum structures of Rigetti, Google, and IBM QPUs, the possible mapping results of the physical connectivity requirements of qubits $a,b,c,d,e$ in Figure~\ref{fig:Triangle_layout1} are demonstrated in Figure~\ref{fig:sym_structs_map}. Notice that the duplications of specific qubits are due to the missing of one connected edge of the triangular layout.

% The proposed step-decreasing V-shaped structure (Stesso) in the next section is a set of standard Toffoli gates, and thus also has the symmetry of the fully connectivity requirement of the triangular structure of standard Toffoli gates in Figure~\ref{fig:Triangle_layout1}. 

As shown in Figure~\ref{fig:layout-rigetti-map} and Figure~\ref{fig:layout-google-map}, the additional one SWAP gate indicated by the red line is required due to the location of the target physical qubit $c$, even though the connectivity number of its neighboring physical qubits is satisfied. For IBM's quantum layout in Figure~\ref{fig:layout-IBM-map}, the additional SWAP gates are mainly required because of the unsatisfied connectivity number of the neighboring physical qubits of the target qubits $c$. When IBM increases the number of neighboring physical connectivity, this will increase the symmetrical logical structures, which means increasing the $n$ for $(n+1)$-bit Toffoli gates.

In Figure~\ref{fig:layout-IBM-map}, the longer red line is due to the location of the target physical qubit $c$, the mapping result is affected by the native gates (like RZ and X gates for single-qubit, CZ gates for two-qubit) and the conventional decomposition method of a standard Toffoli gate. 
% Here, we list two possible ways to decompose a standard Toffoli gate and introduce the different generating physical connectivity requirements.
% On the one hand, one can use Barenco's decomposition of a standard Toffoli gate and further implement the involved elementary gates utilizing the native gates of IBM. This will generate the same full physical connectivity for implementing a standard Toffoli gate in Figure~\ref{fig:Triangle_layout1}. On the other hand, one can also use a different decomposition in GALA-$n$~\cite{galaquantumlibrary} and CALA-$n$~\cite{calaquantumlibrary} libraries to cost-effectively realize a standard Toffoli gate. Compared to the triangular structure, the required physical connectivity of a standard Toffoli gate will be less, i.e., no need for the physical connectivity for two control qubits $a$ and $b$.
The actual physical connectivity requirements of a standard Toffoli gate are generated for the real implementation of a standard Toffoli gate, which is not
only affected by the theoretical decomposition of a standard Toffoli gate. For this reason, our introduced Stesso consists of standard Toffoli gates, with no further focus on the decomposition of standard Toffoli gates.

\subsection{Basic shaped structures}

To effectively decompose $(n+1)$-bit Toffoli gates $\text{C}^n\text{X}$, we introduce different shapes of a set of standard Toffoli gates $\text{C}^2\text{X}$ and X gates using adjustable number of support qubits as declared in Definition~\ref{def:sup_qubit}, named the “step-decreasing structures shaped operators (Stesso)”.
\begin{definition}
\label{def:sup_qubit}
Support qubits $q_s$ are a subset of the ancilla qubits of an $(n+1)$-bit Toffoli gate, where their initial states do not affect the final state of the target qubit $q_t$.
\end{definition}

Here, we define all these basic logical-shaped structures used in Stesso by the number of involved qubits (controls, supports, and target) and the rules they satisfy, such as stated in Definition~\ref{def:backslash-shape}. 
These shapes are symbolically based logical structures, including symbol-shape (like $\backslash$-shape), English-shape (like V-shape), and Cyrillic-shape (like {\foreignlanguage{russian}{И}}-shape). Notice that the V-shape and {\foreignlanguage{russian}{И}}-shape consist of a set of two symbol-shapes of $\backslash$-shape and /-shape, with different numbers of $n$ control qubits.

\begin{definition}
    \label{def:backslash-shape}
    The $\backslash$-shape structure, consisting of $n$ control qubits $q_{c_i}$, $(n - 2)$ support qubits $q_{s_i}$, and one target qubit $q_t$, is a set of standard Toffoli gates satisfying any one of the following four rules, where $n \geq 3$ and $i \geq 1$:
    \begin{itemize}
        \item Control $\rightarrow$ control: connect one control qubit of the $i$-th standard Toffoli gate to the one control qubit of the $(i+1)$-th standard Toffoli gate.
        \item Target $\rightarrow$ target: connect the target qubit of the $i$-th standard Toffoli gate to the target qubit of the $(i+1)$-th standard Toffoli gate.
        \item Control $\rightarrow$ target: connect one control qubit of the $i$-th standard Toffoli gate to the target qubit of the $(i+1)$-th standard Toffoli gate.
        \item Target $\rightarrow$ control: connect the target qubit of the $i$-th standard Toffoli gate to the one control qubit of the $(i+1)$-th standard Toffoli gate.
    \end{itemize}
\end{definition}

In the quantum domain, circuit size and circuit depth are the key metrics for analyzing the complexity requirements to implement a quantum circuit.
For the analysis of the complexity and the usage of a quantum structure, we calculated the circuit size, circuit depth, and number of outputs, and further illustrated the output form of all logical-shaped structures as properties.

In this paper, the circuit size $S$ of a quantum structure (circuit) typically refers to the total number of quantum gates it contains. This is distinct from the circuit depth $D$, which measures the number of sequential vertical layers. A single layer can contain multiple quantum gates as long as they act on different qubits and can be executed in parallel. 
Especially, when each quantum gate contains the qubits of its follow-up gates of a quantum structure, its circuit depth $D$ equals its circuit size $S$, such as described in Property~\ref{pro:bslash-ds}.

\begin{property}
\label{pro:bslash-ds}
The circuit depth of $\backslash$-shape structure is equal to its circuit size, i.e., $D(\backslash) = S(\backslash) = \text{No. C}^2\text{X} = n - 1$.
\end{property}

\begin{figure}[htbp]
    \centering
    \begin{subfigure}[b]{0.15\textwidth}
        \includegraphics[width=0.5\textwidth]{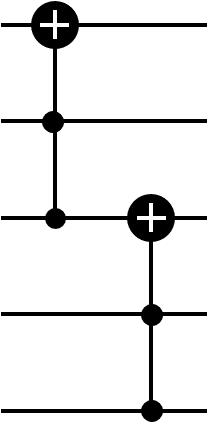}
        \caption{$\backslash$-shape with control $\rightarrow$ target}
        \label{fig:bslash_exp_c2t}
    \end{subfigure}
    %\hfill
     \hspace{0.1\textwidth}
    \begin{subfigure}[b]{0.18\textwidth}
        \includegraphics[width=0.7\textwidth]{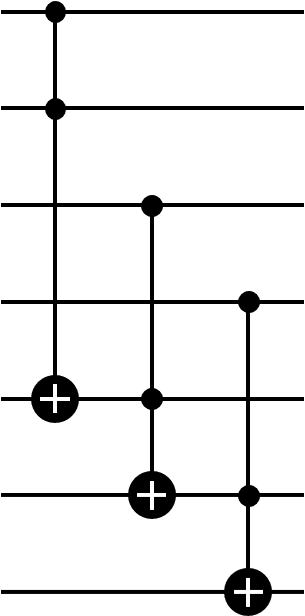}
        \caption{$\backslash$-shape with target $\rightarrow$ control}
        \label{fig:bslash_exp_t2c}
    \end{subfigure}
    \caption{Examples of $\backslash$-shape structures with two satisfied rules: (a) a $\backslash$-shape  (consisting of three controls, one support, and one target qubit;  satisfying the rule of control $\rightarrow$ target qubit) has the same circuit depth and circuit size $D=S= \text{No. C}^2\text{X}=2$, and (b) a $\backslash$-shape (consisting of four controls, two supports, and one target qubit; satisfying the rule of target $\rightarrow$ control qubit) has $D=S= \text{No. C}^2\text{X}=3$.}
    \label{fig:bslash_exp}
\end{figure}

The commonly used $\backslash$-shape structures always satisfy the rule that control goes to the target qubit in Figure~\ref{fig:bslash_exp_c2t}, or the rule that the target goes to one control qubit in Figure~\ref{fig:bslash_exp_t2c}. These two $\backslash$-shape structures in Figure~\ref{fig:bslash_exp} are useful to generate the multiplication form of $(q_{c_i}q_{c_{i+1}} \oplus q_{s_i})(q_{c_i+2}q_{c_{i+3}} \oplus q_{s_{i+1}})$ and $(q_{c_i}q_{c_{i+1}} \oplus q_{s_i})q_{c_{i+2}} \oplus q_{s_{i+1}}$ for different decomposition of $(n+1)$-bit Toffoli gates, as stated in Property~\ref{pro:bslash-out_c2t} ans Property~\ref{pro:bslash-out_t2c} respectively.

\begin{property}
    \label{pro:bslash-out_c2t}
    There are $\frac{n}{2}$ outputs of a $\backslash$-shape structure that follows the rule control $\rightarrow$ target, consisting of one output for the target qubit, and $\frac{n}{2} - 1$ outputs for all support qubits.
    The outputs are in the multiplication form of $(q_{c_i}q_{c_{i+1}} \oplus q_{s_i})$.
\end{property}

\begin{property}
    \label{pro:bslash-out_t2c}
    There are $n - 1$ outputs of a $\backslash$-shape structure that follows the rule target $\rightarrow$ control, consisting of one output for the target qubit, and $n - 2$ outputs for all support qubits.
    The outputs are in the form of $(q_{c_i}q_{c_{i+1}} \oplus q_{s_i})q_{c_{i+2}} \oplus q_{s_{i+1}}$
   .
\end{property}

By adding several X gates into these two $\backslash$-shape structures in Figure~\ref{fig:bslash_exp}, the downstairs(\drawdownstairs{1})-shape structures can be defined in Definition~\ref{def:downstairs-shape}, which makes the reusable of input qubits and generates of negative polarities possible for sub-product terms of $(n+1)$-bit Toffoli gates. Therefore, the number of contained X gates of a downstairs(\drawdownstairs{1})-shape structure is based on the number of negative polarities of sub-product terms $\overline{q_{c_i}q_{c_{i+1}}}$ of $n$ control qubits.

%The details are later explained in the definition of Stesso and its mathematical proof.

\begin{definition}
    \label{def:downstairs-shape}
    The downstairs(\drawdownstairs{1})-shape structure consisting of $n$ control qubits $q_{c_i}$, $(n - 2)$ support qubits $q_{s_i}$, and one target qubit $q_t$ is a set of standard Toffoli gates and Pauli-X (X) gates satisfying any one of the following two rules, where $n \geq 3$:
    \begin{itemize}
        \item Control $\rightarrow$ target: add X gates after the target qubit of a standard Toffoli gate, iff the polarity of its product term of two control qubits is negative.
        \item Target $\rightarrow$ control: add X gates before the target qubit of a standard Toffoli gate, iff the polarity of its product term of two control qubits is negative.
    \end{itemize}
\end{definition}

\begin{property}
\label{pro:dstairs_ds}
    The circuit size of a downstairs(\drawdownstairs{1})-shape structure is based on the number $n_{neg}$ of negative polarities of sub-product terms of $n$ control qubits, i.e., $S(\drawdownstairs{1}) = \text{No. C}^2\text{X} + \text{No. X} = (n - 1) + n_{neg}$, for $ n_{neg} \leq n - 1$.
    The circuit depth of $\drawdownstairs{1}$-shape structure is equal to its circuit size, i.e., $D(\drawdownstairs{1}) = S(\drawdownstairs{1}) = (n - 1) + n_{neg}$
   .
\end{property}

\begin{figure}[htbp]
    \centering
    \begin{subfigure}[b]{0.2\textwidth}
        \includegraphics[width=0.8\textwidth]{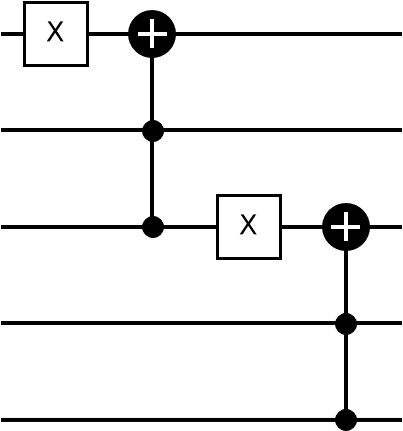}
        \caption{downstairs-shape with control $\rightarrow$ target}
        \label{fig:dstairs_exp_c2t}
    \end{subfigure}
    %\hfill
     \hspace{0.1\textwidth}
    \begin{subfigure}[b]{0.25\textwidth}
        \includegraphics[width=\textwidth]{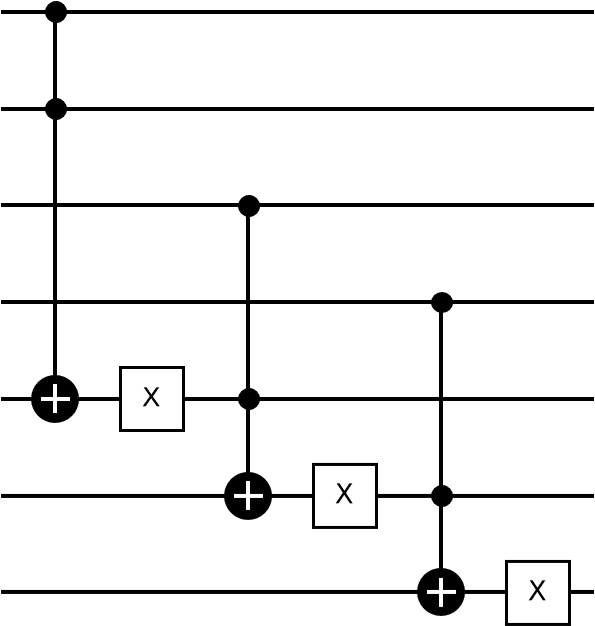}
        \caption{downstairs-shape with target $\rightarrow$ control}
        \label{fig:dstairs_exp_t2c}
    \end{subfigure}
    \caption{Examples of downstairs-shape structures with two satisfied rules: (a) a downstairs-shape (consisting of three controls, one support, and one target qubit; satisfying the rule of control $\rightarrow$ target qubit) has $D= S = \text{No. C}^2\text{X} + \text{No. X}=4$, and (b) a downstairs-shape (consisting of four controls, two supports, and one target qubit; satisfying the rule of target $\rightarrow$ control qubit) has $D=S = \text{No. C}^2\text{X} + \text{No. X}=6$.}
    \label{fig:dstairs_exp}
\end{figure}

For example, the downstairs(\drawdownstairs{1})-shape satisfying the rule of control $\rightarrow$ target shown in Figure~\ref{fig:dstairs_exp_c2t} can be used to generate the multiplication form $(q_{c_2}q_{c_{3}} \oplus \overline{q_{c_1}})({q_{c_4}q_{c_{5}}} \oplus \overline{q_{c_3}}) = q_{c_2}q_{c_{3}}q_{c_4}q_{c_{5}} \oplus \overline{q_{c_1}}({q_{c_4}q_{c_{5}}} \oplus \overline{q_{c_3}})$ in Property~\ref{pro:dstairs_out_c2t}, if the inputs of the circuit from top-to-bottom are all control qubits $q_{c_1},q_{c_2},\cdots,q_{c_5}$. In this way, the inputs of the previous structure can be reused as the support qubits of the current structure, with the additional term $\overline{q_{c_1}}({q_{c_4}q_{c_{5}}} \oplus \overline{q_{c_3}})$ being a step-dereasing downstairs(\drawdownstairs{1})-shape repeatedly.

\begin{property}
    \label{pro:dstairs_out_c2t}
    For a downstairs($\drawdownstairs{1}$)-shape that follows the rule of control $\rightarrow$ target, there are $\frac{n}{2}$ outputs that one for the target qubit and $\frac{n}{2} - 1$ for support qubits. The useful single output of the target is in the multiplication form of $(q_{c_{i}}q_{c_{i+1}} \oplus \overline{q_{s_{i}}})$ iff the product term of the two control qubits $q_{c_{i}},q_{c_{i+1}}$ has negative polarities.
\end{property}

For the downstairs(\drawdownstairs{1})-shape satisfying the rule of target $\rightarrow$ control shown in Figure~\ref{fig:dstairs_exp_t2c}, the added $X$ gates are used to generate the negative polarities of specific sub-product terms of the final states of an $(n+1)$-bit Toffoli gate. If the inputs of the circuit in Figure~\ref{fig:dstairs_exp_t2c} from top-to-bottom are in the order of control qubits $q_{c_1},q_{c_2},q_{c_3},q_{c_4}$, support qubits $q_{s_1},q_{s_2}$, and target qubit $q_t$, then the final state of target qubit is $\overline{\overline{\overline{q_{c_1}q_{c_2}}q_{c_3}}q_{c_4}} \oplus q_t$ as stated in Property~\ref{pro:dstairs_out_t2c}.

\begin{property}
    \label{pro:dstairs_out_t2c}
    For a downstairs($\drawdownstairs{1}$)-shape that follows the rule of target $\rightarrow$ control, there are $n - 1$ outputs that one for the target qubit and $n - 2$ for support qubits. The outputs are in the form of $\overline{\overline{\overline{q_{c_1}q_{c_2}} \cdots} q_{c_n}} \oplus q_t$ iff the sub-product term of the control qubits has negative polarities.
\end{property}

Thus, the downstairs(\drawdownstairs{1})-shape structures in Definition~\ref{def:downstairs-shape} are not only considered as an extension of $\backslash$-shape structures in Definition~\ref{def:backslash-shape} by inserting X gates, but also have construction advantages.
Concretely, such construction advantages of the downstairs(\drawdownstairs{1})-shape structures extend the reconfigurability of $(n+1)$-bit Toffoli gates with the same physical connectivity requirements and polynomial complexity $O(n)$ of circuit depth and circuit size. These structures can reuse the control qubits of $(n+1)$-bit Toffoli gates so that they support an adjustable number of support qubits for Stesso.
Our methodology of using X gates and standard Toffoli gates (in Definition~\ref{def:backslash-shape} and Definition~\ref{def:downstairs-shape}) is an essential key component of designing Stesso.

To maintain the layout-awareness for symmetrical structures (same physical connectivity requirements) and the reconfigurability of $(n+1)$-bit Toffoli gates, the /-shape and upstairs(\drawupstairs{1})-shape structures are constructed as the reversed $\backslash$-shape and downstairs(\drawdownstairs{1})-shape structures, as stated in Definition~\ref{def:slash-shape} and Definition~\ref{def:upstairs-shape}, respectively.

\begin{definition}
    \label{def:slash-shape}
    The /-shape structure consisting of $n$ control qubits $q_{c_i}$, $(n - 2)$ support qubits $q_{s_i}$, and one target qubit $q_t$ is a set of standard Toffoli gates satisfying any one of the following four rules, where $n \geq 3$ and $i \geq 1$:
    \begin{itemize}
        \item Control $\rightarrow$ control: connect one control qubit of the $i$-th standard Toffoli gate to the one control qubit of the $(i+1)$-th standard Toffoli gate.
        \item Target $\rightarrow$ target: connect the target qubit of the $i$-th standard Toffoli gate to the target qubit of the $(i+1)$-th standard Toffoli gate.
        \item Control $\rightarrow$ target: connect one control qubit of the $i$-th standard Toffoli gate to the target qubit of the $(i+1)$-th standard Toffoli gate.
        \item Target $\rightarrow$ control: connect the target qubit of the $i$-th standard Toffoli gate to the one control qubit of the $(i+1)$-th standard Toffoli gate.
    \end{itemize}
\end{definition}

% \begin{property}
%     \label{pro:slash-ds}
%     The circuit depth of /-shape structure is equal to its circuit size, i.e., $D(/) = S(/) = \text{No. C}^2\text{X} = n - 1$
%    .
% \end{property}

% \begin{property}
%     \label{pro:slash-out_c2t}
%     There are $n - 1$ outputs of a /-shape structure that follows the rule control $\rightarrow$ target, consisting of one output for the target qubit, and $n - 2$ outputs for all support qubits.
%     The outputs are in the form of 
%     $(q_{c_i}q_{c_{i+1}} \oplus q_{s_i})$
%    .
% \end{property}

% \begin{property}
%     \label{pro:slash-out_t2c}
%     There are $\frac{n}{2}$ outputs of a /-shape structure that follows the rule target $\rightarrow$ control, consisting of one output for the target qubit, and $\frac{n}{2} - 1$ outputs for all support qubits.
%     The outputs are in the multiplication form of $(q_{c_i}q_{c_{i+1}} \oplus q_{s_i})$
%    .
% \end{property}

Notice that the two examples of /-shape structures that meet the rules between control and target are the different reversed $\backslash$-shape structures of opposite rules. That is, the /-shape structures with rule control $\rightarrow$ control or rule target $\rightarrow$ target are the reversed $\backslash$-shape structures of the same rule.
The /-shape structure with rule control $\rightarrow$ target is the reversed $\backslash$-shape structure of the rule target $\rightarrow$ control.
The \-shape structure with rule target $\rightarrow$ control is the reversed $\backslash$-shape structure of the rule control $\rightarrow$ target.

% \begin{figure}[htbp]
%     \centering
%     \begin{subfigure}[b]{0.18\textwidth}
%         \includegraphics[width=0.7\textwidth]{figs/slash-c2t.png}
%         \caption{/-shape with control $\rightarrow$ target}
%         \label{fig:slash_exp_c2t}
%     \end{subfigure}
%     %\hfill
%      \hspace{0.1\textwidth}
%     \begin{subfigure}[b]{0.15\textwidth}
%         \includegraphics[width=0.5\textwidth]{figs/slash-t2c.png}
%         \caption{/-shape with target $\rightarrow$ control}
%         \label{fig:slash_exp_t2c}
%     \end{subfigure}
%     \caption{Examples of /-shape structures with two satisfied rules: (a) a /-shape (consisting of four controls, two supports, and one target qubit; satisfying the rule of control $\rightarrow$ target qubit) has the same circuit depth and circuit size $D=S= \text{No. C}^2\text{X}=3$, and (b) a /-shape (consisting of three controls, one support, and one target qubit; satisfying the rule of target $\rightarrow$ control qubit) has $D=S= \text{No. C}^2\text{X}=2$.}
%     \label{fig:slash_exp}
% \end{figure}

For the upstairs(\drawupstairs{1})-shape structures stated in Definition~\ref{def:upstairs-shape} with the rules between control and target qubits, they are also the reversed downstairs(\drawdownstairs{1})-shape structures stated in Definition~\ref{def:downstairs-shape} that satisfy the opposite of the rules.

\begin{definition}
    \label{def:upstairs-shape}
    The upstairs(\drawupstairs{1})-shape structure consisting of $n$ control qubits $q_{c_i}$, $(n - 2)$ support qubits $q_{s_i}$, and one target qubit $q_t$ is a set of standard Toffoli gates and Pauli-X (X) gates satisfying any one of the following two rules, where $n \geq 3$:
    \begin{itemize}
        \item Control $\rightarrow$ target: add X gates before the target qubit of a standard Toffoli gate, iff the polarity of its product term of two control qubits is negative.
        \item Target $\rightarrow$ control: add X gates after the target qubit of a standard Toffoli gate, iff the polarity of its product term of two control qubits is negative.
    \end{itemize}
\end{definition}

% \begin{property}
%     \label{pro:ustair_ds}
%     The circuit size of an upstairs(\drawupstairs{1})-shape structure is based on the number $n_{neg}$ of negative polarities of sub-product terms of $n$ control qubits.
%     The circuit depth of $\drawupstairs{1}$-shape structure is equal to its circuit size, i.e., $D(\drawupstairs{1}) = S(\drawupstairs{1}) = \text{No. C}^2\text{X} + \text{No. X} = (n - 1) + n_{neg}$ for $n_{neg} \leq n - 1$
%    .
% \end{property}

% \begin{property}
%     \label{pro:ustairs-out_c2t}
%     There are $n - 1$ outputs of an upstairs(\drawupstairs{1})-shape structure that follows the rule control $\rightarrow$ target, consisting of one output for the target qubit, and $n - 2$ outputs for all support qubits.
%     The outputs are in the form of     $(\overline{q_{c_i}q_{c_{i+1}}} \oplus q_{s_i})$ iff the product term of the two control qubits $q_{c_{i}},q_{c_{i+1}}$ has negative polarities
%    .
% \end{property}

% \begin{property}
%     \label{pro:ustairs-out_t2c}
%     There are $\frac{n}{2}$ outputs of an upstairs(\drawupstairs{1})-shape structure that follows the rule target $\rightarrow$ control, consisting of one output for the target qubit, and $\frac{n}{2} - 1$ outputs for all support qubits.
%     The outputs are in the multiplication form of $(\overline{q_{c_i}q_{c_{i+1}}} \oplus q_{s_i})$ iff the product term of the control qubits has negative polarities
%    .
% \end{property}

\begin{figure}[htbp]
    \centering
    \begin{subfigure}[b]{0.25\textwidth}
        \includegraphics[width=\textwidth]{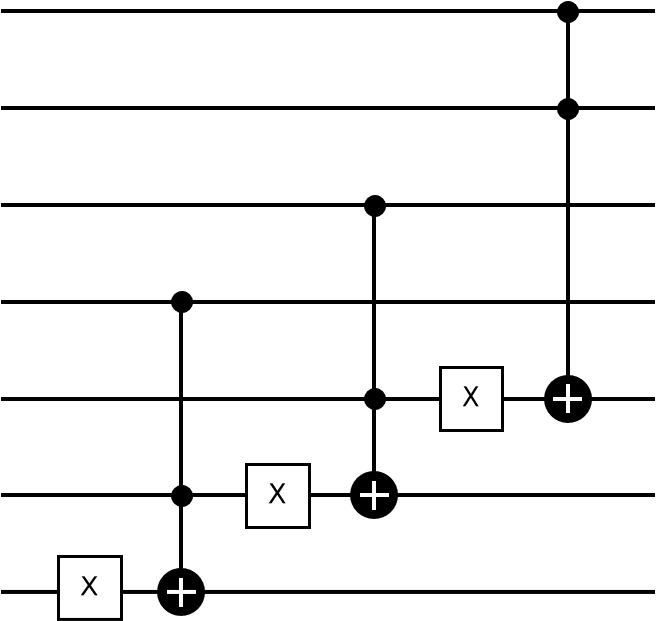}
        \caption{upstairs-shape with control $\rightarrow$ target}
        \label{fig:ustairs_exp_c2t}
    \end{subfigure}
    %\hfill
     \hspace{0.1\textwidth}
    \begin{subfigure}[b]{0.2\textwidth}
        \includegraphics[width=0.8\textwidth]{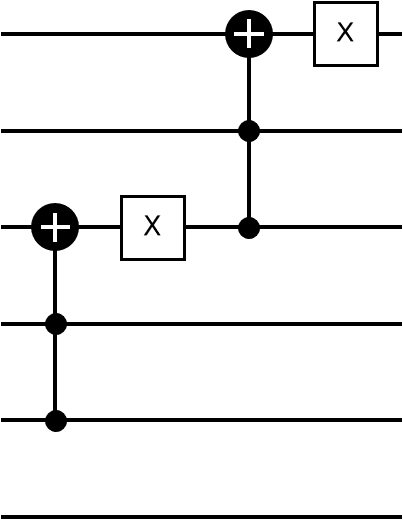}
        \caption{upstairs-shape with target $\rightarrow$ control}
        \label{fig:ustairs_exp_t2c}
    \end{subfigure}
    \caption{Examples of upstairs-shape structures with two satisfied rules: (a) an upstairs-shape (consisting of four controls, two supports, and one target qubit; satisfying the rule of control $\rightarrow$ target qubit) has $D= S = \text{No. C}^2\text{X} + \text{No. X}=6$, and (b) an upstairs-shape (consisting of three controls, one support, and one target qubit; satisfying the rule of target $\rightarrow$ control qubit) has $D=S = \text{No. C}^2\text{X} + \text{No. X}=4$.}
    \label{fig:ustairs_exp}
\end{figure}

Therefore, Figure~\ref{fig:ustairs_exp_c2t} and Figure~\ref{fig:ustairs_exp_t2c} demonstrate the reversed circuits shown in Figure~\ref{fig:dstairs_exp_t2c} and Figure~\ref{fig:dstairs_exp_c2t}, respectively.
For instance, the upstairs(\drawupstairs{1})-shape structure, which follows the rule of control $\rightarrow$ target, can be used to generate terms of $(q_{c_4}q_{s_{2}} \oplus \overline{q_{t}})$, $(q_{c_3}q_{s_{1}} \oplus \overline{q_{s_2}})$ and $(q_{c_1}q_{c_{2}} \oplus \overline{q_{s_1}})$, if the inputs of the circuit in Figure~\ref{fig:ustairs_exp_c2t} from top-to-bottom are in the order of control qubits $q_{c_1},q_{c_2},q_{c_3},q_{c_4}$, support qubits $q_{s_1},q_{s_2}$, and target qubit $q_t$.
If a user composes the circuits in Figure~\ref{fig:dstairs_exp_t2c} and a step-decreasing structure similar with the structure in Figure~\ref{fig:ustairs_exp_c2t}, then the final state of target qubit is $\overline{\overline{\overline{q_{c_1}q_{c2}}q_{c_3}}q_{c_4}} \oplus q_t$, while the final states of other qubits remain unchanged. Our reconfigurability of Stesso is introduced in the ``Step-decreasing structures shaped operators" section~\ref{sec:SteSSO}, following a similar methodology.

% \begin{equation}
%     c_1c_2 + c_3 + c_4
%     \text{De Morgan's principle, use or operation +}
% \end{equation}

Different from the above-mentioned symbol-shape structures and their expansions, including $\backslash$-shape, downstairs(\drawdownstairs{1})-shape, /-shape, and upstairs(\drawupstairs{1})-shape structures, other basic shaped structures such as V-shape (English-shape) in Definition~\ref{def:v-shape} and {\foreignlanguage{russian}{И}}-shape (Cyrillic-shape) in Definition~\ref{def:rev-N-shape} can be decomposed into a set of symbol-shapes. These shaped structures are defined as basic shaped structures for less consideration of the composition ability of support qubits, with only one useful output in Property~\ref{pro:vshape-out} and Property~\ref{pro:rev_N_out} respectively. In contrast, other symbol-shape structures always have multiple outputs, which are the supports and the target qubit. 

\begin{definition}
    \label{def:v-shape}
    The V-shape structure consisting of $n$ control qubits $q_{c_i}$, $(n - 2)$ support qubits $q_{s_i}$, and one target qubit $q_t$ is the composed circuit of both $\backslash$-shape structure for $n$ control qubits (satisfying the rule target $\rightarrow$ control) and /-shape structure for $(n-1)$ control qubits (satisfying the rule control $\rightarrow$ target), where $n \geq 3$.
    % \prod_{i=1}^{n-2}{q_{c_i}q_{c_{i+1} \oplus q_{s_{i}}}}=
\end{definition}

\begin{property}
    \label{pro:vshape-ds}
    The circuit depth $D$ of a V-shape structure is equal to its circuit size $S$, i.e., $D(V) = S(V) = S(\backslash)+S(/) = \text{No. C}^2\text{X} = n - 1 + n - 2 = 2n - 3$
   .
\end{property}

\begin{property}
    \label{pro:vshape-out}
    The single output of a V-shape structure is in the form of $(\cdots((q_{c_1}q_{c_2} \oplus q_{s_1})q_{c_3} \oplus q_{s_2})\cdots q_{c_{n-1}} \oplus q_{s_{n-2}})q_{c_n} \oplus q_t = q_t \oplus \prod_{i = 1}^{n}q_{c_i} \oplus (\cdots(q_{s_1}q_{c_3} \oplus q_{s_2})\cdots q_{c_{n-1}} \oplus q_{s_{n-2}})q_{c_n}$
   .
\end{property}

\begin{definition}
    \label{def:rev-N-shape}
    The {\foreignlanguage{russian}{И}}-shape structure consisting of $n$ control qubits $q_{c_i}$, $(n - 2)$ support qubits $q_{s_i}$, and one target qubit $q_t$ is the composed circuit of three $\backslash$-shape structure for $n$ control qubits (satisfying the rule target $\rightarrow$ control), /-shape structure for $(n-1)$ control qubits (satisfying the rule control $\rightarrow$ target) and $\backslash$-shape for $(n-1)$ control qubits (satisfying the rule target $\rightarrow$ control), where $n \geq 3$
   .
\end{definition}

\begin{property}
    \label{pro:rev_N_ds}
    The circuit depth $D$ of a {\foreignlanguage{russian}{И}}-shape structure is equal to its circuit size $S$, i.e., $D(\text{\foreignlanguage{russian}{И}}) = S(\text{\foreignlanguage{russian}{И}}) = S(V) + S(\backslash) = S(\backslash)+S(/) + S(\backslash) = \text{No. C}^2\text{X} = 2n -3 + n - 2 = 3n - 5$
   .
\end{property}

\begin{property}
    \label{pro:rev_N_out}
    There are $n-2$ outputs of a {\foreignlanguage{russian}{И}}-shape structure when $n \geq 3$. That is, a useful output $q_t \oplus \prod_{i = 1}^{n}q_{c_i}$ for an $(n+1)$-bit Toffoli gate, and $n-3$ auxiliary outputs of support qubits. When there are three control qubits $(n=3)$ of a {\foreignlanguage{russian}{И}}-shape structure, only one single output $q_t \oplus q_{c_1}q_{c_2}q_{c_3}$ exists.
   .
\end{property}

\begin{figure}[htbp]
    \centering
    \begin{subfigure}[b]{0.4\textwidth}
        \includegraphics[width=\textwidth]{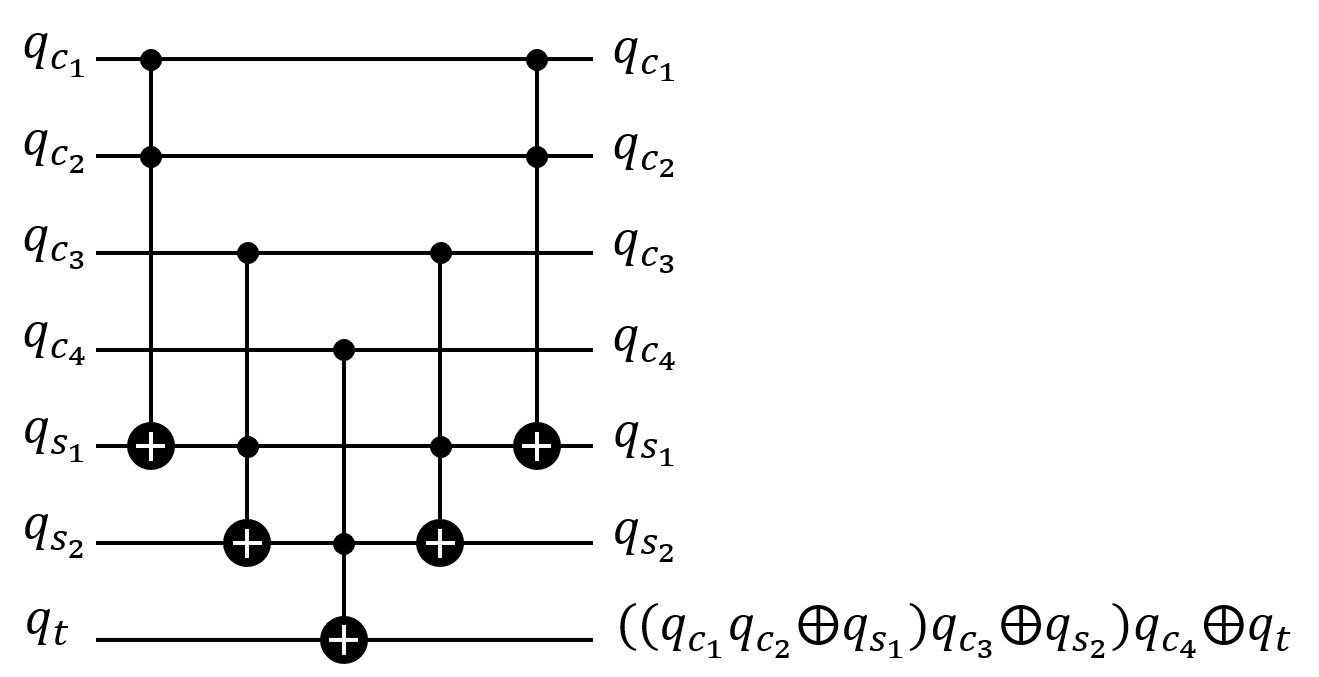}
        \caption{V-shape}
        \label{fig:v_exp}
    \end{subfigure}
    %\hfill
     \hspace{0.1\textwidth}
    \begin{subfigure}[b]{0.2\textwidth}
        \includegraphics[width=\textwidth]{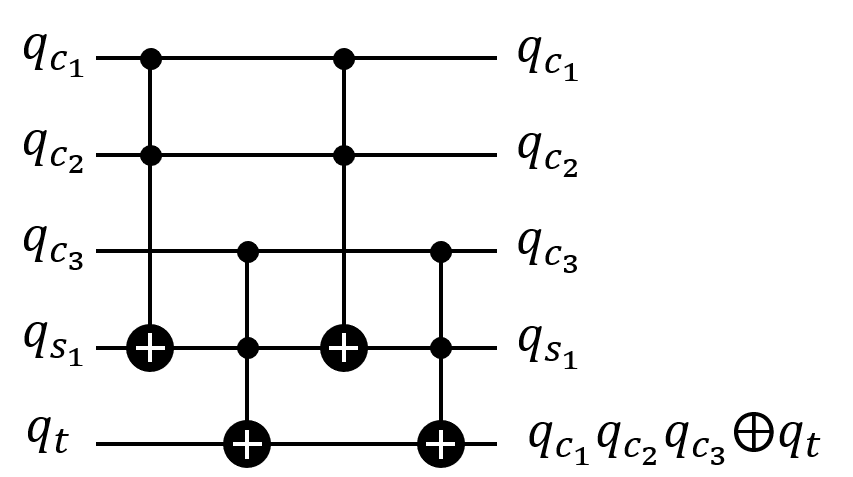}
        \caption{{\foreignlanguage{russian}{И}}-shape}
        \label{fig:rev_N_exp}
    \end{subfigure}
    \caption{Examples of a V-shape structure and {\foreignlanguage{russian}{И}}-shape structure with a single output: (a) a V-shape structure (consisting of four control qubits, two support qubits, and one target qubit) has $D= S =5$, and (b) a {\foreignlanguage{russian}{И}}-shape structure (consisting of three control qubits, one support qubit, and one target qubit) has $D=S = 4$.}
    \label{fig:v_N_exp}
\end{figure}

For example, there is only one useful output for each target qubit $q_t$ utilizing a V-shape structure and a {\foreignlanguage{russian}{И}}-shape structure in Figure~\ref{fig:v_N_exp}, respectively. The output $q_t \oplus q_{c_1}q_{c_2}q_{c_3}q_{c_4} \oplus q_{s_1}q_{c_3}q_{c_4} \oplus q_{s_2}q_{c_4}$ in Figure~\ref{fig:v_exp} can considered as a composition circuit of a $(4+1)$-bit Toffoli gate and a step-decreasing V-shape structure (consisting of three controls $q_{s_1}, q_{c_3}, q_{c_4}$, one support $q_{s_2}$, and one target). The output in Figure~\ref{fig:rev_N_exp}, which is obtained by a {\foreignlanguage{russian}{И}}-shape structure (consisting of three controls, one support, and one target), can be directly considered as the decomposition of an $(3+1)$-bit Toffoli gate.
Meanwhile, the states of all controls and supports of these two examples remain unchanged.

\subsection{Step-decreasing structures shaped operators}
\label{sec:SteSSO}

Based on the above-mentioned basic logical-shaped structures, our ``step-decreasing structures shaped operators (Stesso)" can handle different polarities of sub-product terms of $(n+1)$-bit Toffoli gates by utilizing standard Toffoli gates ($\text{C}^2\text{X}$) and Pauli-X (X) gates. 
To extend the reconfigurability of the conceptual design for representing $(n+1)$-bit Toffoli gates, we extend the set of all contained gates (not only containing $\text{C}^2\text{X}$ and X gates) by adding constrained unitary gates (like CX gates) to the Stesso.
There are three types of Stesso, which can handle the positive polarities, mixed polarities, and mixed operators (including Negation, AND, OR, and XOR operators) of control qubits for $(n+1)$-bit Toffoli gates, which are named as ``PP-Stesso", ``MP-Stesso", and ``G-Stesso" respectively.
All three types of Stesso are composed of two similar composed shape structures in Definition~\ref{def:composed_shapes}, with two different control qubits $n$ and $n-1$.

\begin{definition}
    \label{def:composed_shapes}
    The composed shaped structures are sequentially composed by repetitions $r$ and transitions $t$ of the basic shaped structures (including symbol-shapes, English-shapes, and Cyrillic-shapes) following the sequence $Seq = \{ (r_1$, $f_{r_1}$, $\backslash)$, $(t_1,f_{t_1}, \backslash)$, $(r_2,f_{r_2},\drawdownstairs{1})$, $(t_2,f_{t_2}, \drawdownstairs{1})$, $(r_3, f_{r_3}, V)$, $(t_3, f_{t_3},V)$, $(r_4, f_{r_4},\text{\foreignlanguage{russian}{И}})$, $(t_4,f_{t_4}$, $\text{\foreignlanguage{russian}{И}})$, $(r_5,f_{r_5},\drawupstairs{1})$, $(t_5,f_{t_5},\drawupstairs{1})$, $(r_6, f_{r_6},/)\}$ that has $\Sigma_{i=1}^{6}{r_i}+\Sigma_{i=1}^{5}{t_i} \geq 1$, where $r_i \in \mathbb{N}$, $t_i \in \{ 0,1\}$, $f_i$ is a function to obtain the specific tetrad $(\{q_c\},\{q_s\},\{q_t\},rule)$ of control set, support set, target set, and satisfied rule for a basic shaped structure
   .
\end{definition}

As stated in Definition~\ref{def:composed_shapes}, each element of the combination sequence $Seq$ of a composed shaped structures is a triple of repetition number $r_i$ (or transition number $t_i$), function $f_{r_i}$ (or $f_{t_i}$) to obtain tetrad $(\{q_c\},\{q_s\},\{q_t\},rule)$, and a basic shaped structure. For instance, a triple of $(r_1,f_{r_1},\backslash)$ means the composition of $r_1$ $\backslash$-shaped structures for different controls $q_c$, supports $q_s$, target $q_t$, and satisfied rule of each repetition. 

% \begin{figure}[htbp]
%     \centering
%     \includegraphics[width=0.8\textwidth]{figs/Stesso_PP.png}
%     \caption{Combination sequence of the composed shaped structures used in Stesso. The $r_i$ and $t_i$ represent the repetitions and transitions of different-shaped structures, respectively. And the range of transition numbers belong to $\{0,1\}$.}
%     \label{fig:Stesso_PP}
% \end{figure}

Obviously, if there is none repetitions and transitions of basic shaped structures, the composed shaped structure will be a empty circuit and meaningless.
%If there is only one repetition or transition in total of the combination sequence for the basic shaped structures, then the composed shaped structure will be just a single basic shaped structure.
So only when the sum of all repetition and transitions is not zero, i.e., $\Sigma_{i=1}^{6}{r_i}+\Sigma_{i=1}^{5}{t_i} \geq 1$, the composed shaped structures based on different basic shaped structures are not empty circuits and meaningful. 

Notice that the output of a composed shaped structure is based on the combination sequence of all basic shaped structures. If one want to obtain a possible output of a composed shaped structure, there may exists several possible different composed shaped structures to generate the same output. 

% \begin{property}
% \label{pro:composed_exist}
%     When the outputs is able to obtained from a specific composed shaped structure, then there at least exists one combination sequences of composed shaped structures to obtain the same outputs.
% \end{property}

For example,
the two different composed shaped structure of the combination sequences $\{(1,(\{q_{c_1},q_{c_2},q_{c_3}\}, \{\overline{q_{s_1}}\}, \{q_t\}, target \rightarrow control), \drawdownstairs{1})$,  $(1,(\{q_{c_1},q_{c_2}\}, \{$ $\}$,$ \{\overline{q_{s_1}}\}$, $ control \rightarrow target),\drawupstairs{1})\}$ and $\{ (1, (\{q_{c_1},q_{c_2},q_{c_3}\}, \{q_{s_1}\}, \{q_t\}$, $ V)\}$ have the same output $q_{c_1}q_{c_2}q_{c_3} \oplus q_t$ for the target qubit with the states of other qubits remain unchanged, where $\overline{q_{s_1}}$ means add X gates related to qubit $q_{s_1}$ according to definitions of downstairs-shape structures (Definition~\ref{def:downstairs-shape}) and upstairs-shape structures (Definition~\ref{def:upstairs-shape}).

Now, based on the definition of the composed shaped structures, the step-decreasing structure operators with positive polarities of $n$ controls is defined in Definition~\ref{def:Stesso-PP}.

\begin{definition}
    \label{def:Stesso-PP}
    The step-decreasing
structures shaped operators are a conceptual design for representing an $(n+1)$-bit Toffoli gate with positive polarities of all $n$ control qubits, such a design is named ``PP-Stesso". The PP-Stesso is a composed circuit of two step-decreasing composed shaped structures, including symbol-shapes, English-shapes, and Cyrillic-shapes
   .
\end{definition}

In this paper, two step-decreasing composed shaped structures means that if the first composed shaped structure involved maximum $n$ control qubits $q_{c_1}, \cdots, q_{c_n}$, then the second composed shaped structure involved maximum $(n - 2)$ control qubits $q_{c_3}, \cdots, q_{c_n}$ with gates related first two controls $q_{c_1}$ and $q_{c_2}$ removed.

According to the Definition~\ref{def:Stesso-PP}, there exists the similar property for the same possible outputs of PP-Stesso, as illustrated in Property~\ref{pro:stesso_pp_exist}.

\begin{property}
\label{pro:stesso_pp_exist}
    When the outputs is able to obtained from a specific PP-Stesso, then there at least exists one PP-Stesso consisting of different composed shaped structures to obtain the same outputs.
\end{property}

Since there may exists more than one PP-Stesso to generate same outputs in Property~\ref{pro:stesso_pp_exist}, the circuit depth and circuit size of different PP-Stesso are based on the calculation of the combination sequence of their corresponding composed shaped structures.

% \begin{property}
%     \label{pro:stesso_pp_ds}
%     If the combination sequence $Seq$ of the composed shaped structures used in a PP-Stesso is $\{ (r_1, f_{r_1},\backslash)$, $(t_1,f_{t_1}, \backslash)$, $(r_2,f_{r_2},\drawdownstairs{1})$, $(t_2,f_{t_2}, \drawdownstairs{1})$, $(r_3, f_{r_3}$, $ V)$, $(t_3, f_{t_3}, V)$, $(r_4, f_{r_4},\text{\foreignlanguage{russian}{И}})$, $(t_4,f_{t_4}$, $\text{\foreignlanguage{russian}{И}})$, $(r_5,f_{r_5},\drawupstairs{1})$, $(t_5,f_{t_5},\drawupstairs{1})$, $(r_6, f_{r_6}$, $/)\}$, then the circuit depth is $2(\Sigma_{i=1}^{6}{r_i D(shape(f_{r_i}))} + \Sigma_{i=1}^{5}{t_i D(shape(f_{t_i})) - 1)}$. The circuit size is $2(\Sigma_{i=1}^{6}{r_i S(shape(f_{r_i}))} + \Sigma_{i=1}^{5}{t_i S(shape(f_{t_i})) - 1)}$.
% \end{property}

Here we introduce three possible PP-Stessos based on different combination sequence $Seq_1, Seq_2, Seq_3$ of composed shaped structures for an $(n+1)$-bit Toffoli gate, where $n \geq 3$.
Notice that the first PP-Stesso based on $Seq_1$ in Eq.~(\ref{eq:seq_v}) is the same structure proposed of Barenco's paper \cite{barenco1995elementary}, while the other two PP-Stessos are our newly introduced decomposition structure. The first PP-Stesso based on $Seq_1$ is introduced just for the comparison of circuit depth and circuit size with the other two PP-Stessos.

\begin{equation}
    \label{eq:seq_v}
    Seq_1 = \{ (1, (\{q_{c_i} | i \in [ 1, n]\}, \{q_{s_i}| i \in [ 1, n - 2]\}, \{q_t\}, ~),  V)\}
\end{equation}

For the second PP-Stesso based on $Seq_2$ in Eq.~\ref{eq:seq_wave}, the composed shaped structure is composed by one $\backslash$-shape structure, one downstairs(\drawdownstairs{1})-shape structure, two step-decreasing V-shape structures, one upstairs(\drawupstairs{1})-shape structure and one /-shape structure.

%In the Eq.~(\ref{eq:seq_wave}), number $n$ and $n_s$ represents the number of controls and support qubits respectively.
According to the $Seq_2$ in Eq.~(\ref{eq:seq_wave}), the output $\prod_{i = 1}^{n}q_{c_i} \oplus q_t$ of an $(n+1)$-bit Toffoli gate with positive polarities is obtained using the properties of $q_{c_i} \overline{q_{c_i}} = \ket{0}$, $q_{c_i}{q_{c_i}} = {q_{c_i}}$, and $q_{c_i}q_{c_{i+1}} \oplus \ket{0} = q_{c_i}q_{c_{i+1}}$. The X gates in this specific PP-Stesso are used to support the reuse of control qubits, so that the number $n_s$ of all support qubits becomes adjustable in range of $1 \leq n_s \leq n - 2$.
The number $k=\lfloor\frac{n - n_s - 1}{2}\rfloor$ in the $Seq_2$ equals to $\frac{n - n_s - 1}{2}$ when $n - n_s - 1$ is a even number, otherwise equals to $\frac{n - n_s - 2}{2}$. % when $n - n_s - 1$ is a odd number.
The main composition idea of the combination sequence $Seq_2$ for decomposition of an $(n+1)$-bit Toffoli gate can be divided into four steps: (i) firstly generate a sub-product term of $\prod_{i=1}^{n_s + 1}q_{c_i}$ using the $\backslash$-shape structure, (ii) secondly generate $k$ multiplication sub-product form of two controls qubits of ${q_{c_i}q_{c_{i+1}}}$ using the downstairs(\drawdownstairs{1})-shape structure, (iii) generate a $(k+1)+1$-bit or $(k+2)+1$-bit Toffoli gate using two step-decreasing V-shape structures according to whether $n-n_s - 1$ is even number or not, and (iv) finally reversed the two steps (ii) and (i) to make the initial states of all controls and supports remain unchanged.

\begin{equation}
    \label{eq:seq_wave}
    \begin{aligned}
        Seq_2 = 
         \{& 
        (1, (\{q_{c_i} |
        i \in [ 1, n_s + 1] \}, \{q_{s_i}| i \in [ 1, n_s - 1]\}, \{q_{s_{n_s}}\},t \rightarrow c),  \backslash),
        \\&
        (1,(\{q_{c_i} |
        i \in [ n_s + 2, n] \}, \{\overline{q_{c_i}}| i \in [ 1, k - 1]\},
        %\\& \qquad
        \{\overline{q_{c_k}}\}, c \rightarrow t), \drawdownstairs{1}),
        \\&
        (1, (
        \{q_{c_i}| i \in set \}
        %\\& \qquad
        \cup
        \{ q_{s_{n_s}}\} ,
        \\& \qquad
        \{q_{c_i} | i \in [k+1, n], i \notin set \} \cup \{ q_{s_i} | i \in [1,n_s - 1]\},\{q_t\},~), V),
        \\&
        (1, (
        \{q_{c_i}| i \in set \cup \{k+1\}, i \notin [1,2] \}
        %\\& \qquad
        \cup
        \{ q_{s_{n_s}}\} ,
        \\& \qquad
        \{q_{c_i} | i \in [k+2, n], i \notin set \} \cup \{ q_{s_i} | i \in [1,n_s - 1]\},\{q_t\},~), V),
        \\&
        (1,(\{q_{c_i} |
        i \in [ n_s + 2, n] \}, \{\overline{q_{c_i}}| i \in [ 1, k - 1]\},
        %\\& \qquad
        \{\overline{q_{c_k}}\}, t \rightarrow c), \drawupstairs{1}),
        \\& 
        (1, (\{q_{c_i} |
        i \in [ 1, n_s + 1] \}, \{q_{s_i}| i \in [ 1, n_s - 1]\}, \{q_{s_{n_s}}\},c \rightarrow t),  /)
        \},
        \\& 
         k = \lfloor\frac{n - n_s - 1}{2}\rfloor,
        set = [ 1, k] \cup \{ n | n_s + 1 + 2k = n \}
    \end{aligned}
\end{equation}

Differ from the second PP-Stesso by directly decompose the product term $\prod_{i=1}^{n}q_{c_i}$ into a big sub-product term $\prod_{i=1}^{n_s + 1}q_{c_i}$ according to the number of support qubits and several left small sub-product terms of two controls $q_{c_i}q_{c_{i+1}}$, the length of all decomposed sub-product terms of the third PP-Stesso based on $Seq_3$ in Eq.~(\ref{eq:seq_stair}) is a arithmetic progression of $\{2,3,\cdots,k,k+1\}$. However, the final length of the sub-product term can be within the range of $[1,k+1]$ to accommodate all values of total control qubits. Therefore, the number $m$ in Eq.~(\ref{eq:seq_stair}) represents that there are $(m+1)$ control qubits left for the final decomposed sub-product term.

% The main composition idea of the combination sequence $Seq_3$ for the decomposition of an $(n+1)$-bit Toffoli gate can also be divided into four steps: (i) firstly generate a sub-product term of $q_{c_1}q_{c_2}$ using one $\backslash$-shape structure, (ii) recursively generate all sub-product terms of $\prod_{i=\frac{j(j+1)}{2}}^{\frac{(j+1)(j+2)}{2} - 1}{q_{c_i}}$ for $2 \leq j < k$ and final decomposed sub-product term of $\prod_{i=\frac{k(k+1)}{2}}^{\frac{k(k+1)}{2}+m}{q_{c_i}}$ for $0 \leq m \leq k$ using $k - 1$ downstairs-shape structures, where $2 \leq k$, (iii) generate a $k+1$-bit Toffoli gate that all control qubits are the generated sub-product terms by (i) and (ii) using two step-decreasing V-shaped structures, and (iv) finally reversed steps (ii) and (i). 
% In this way, the number of reusable control qubits increases so that there is only one support qubit required to represent an $(n+1)$-bit Toffoli gate.

\begin{equation}
    \label{eq:seq_stair}
    \begin{aligned}
        Seq_3 = 
         \{& 
        (1, (\{q_{c_i} | i \in [ 1, 2] \}, \{\}, \{q_{s_1}\},t \rightarrow c),  \backslash),
        \\&
        (1,(\{q_{c_i} |
        i \in [3, 5] \}, \{\overline{q_{c_1}}\},
        \{\overline{q_{c_2}}\}, t \rightarrow c), \drawdownstairs{1}),
        % \\&
        % (1,(\{q_{c_i} |
        % i \in [6, 9] \}, \{\overline{q_{c_i}}| i \in [3,4] \},
        % \{\overline{q_{c_5}}\}, t \rightarrow c), \drawdownstairs{1}),
        \\& \cdots \qquad \cdots \qquad \cdots
        \\& (1,(\{q_{c_i} |
        i \in [\frac{k(k+1)}{2}, n],  n = \frac{k(k+1)}{2} + m \}, 
        \\& \qquad
        \{\overline{q_{c_i}} |
        i \in [\frac{k(k-1)}{2}, \frac{k(k-1)}{2} + m - 2] \},
        \{\overline{q_{c_{\frac{k(k-1)}{2} + m - 1}}}\},
        \\& \qquad
        t \rightarrow c), \drawdownstairs{1}),
        \\&
        (1, (
        \{q_{c_i}| i \in set \}
        %\\& \qquad
        \cup
        \{ q_{s_1}\} ,
        %\\& \qquad
        \{q_{c_i} | i \in [1, n], i \notin set \},\{q_t\},~), V),
        \\&
        (1, (
        \{q_{c_i}| i \in set \cup \{1\}, i \notin \{2\} \} ,
        %\\& \qquad
        \{q_{c_i} | i \in [2, n], i \notin set \},\{q_t\},~), V),
        \\& (1,(\{q_{c_i} |
        i \in [\frac{k(k+1)}{2}, n],  n = \frac{k(k+1)}{2} + m \}, 
        \\& \qquad
        \{\overline{q_{c_i}} |
        i \in [\frac{k(k-1)}{2}, \frac{k(k-1)}{2} + m - 2] \},
        \{\overline{q_{c_{\frac{k(k-1)}{2} + m - 1}}}\},
        \\& \qquad
        t \rightarrow c), \drawupstairs{1}),
        \\& \cdots \qquad \cdots \qquad \cdots
        % \\&
        % (1,(\{q_{c_i} |
        % i \in [6, 9] \}, \{\overline{q_{c_i}}| i \in [3,4] \},
        % \{\overline{q_{c_5}}\}, t \rightarrow c), \drawupstairs{1}),
        \\&
        (1,(\{q_{c_i} |
        i \in [3, 5] \}, \{\overline{q_{c_1}}\},
        \{\overline{q_{c_2}}\}, t \rightarrow c), \drawupstairs{1}),
        \\&
        (1, (\{q_{c_i} | i \in [ 1, 2] \}, \{\}, \{q_{s_1}\},t \rightarrow c),  /)\},
        \\& 
         0 \leq m \leq k, 2 \leq k,
        set = \{ i | i = \frac{k(k+1)}{2} - 1, 2 \leq k\} \cup \{ n \}
    \end{aligned}
\end{equation}

Then, we compare these three different PP-Stessos based on the combination sequences $Seq_1, Seq_2, Seq_3$ by calculating the number of involved qubits and the number of quantum gates (circuit size) in Table~\ref{tab:cost_stesso_pp}. 

\begin{table}[h!]
    \centering
    \caption{Quantum cost of different PP-Stessos for an $(n+1)$-bit Toffoli gate based on the combination sequences $Seq_1, Seq_2, Seq_3$ of the composed shaped structures.}
    \begin{tabular}{p{2cm}|cc|ccc}
    \hline
    {Combination sequences} & \multicolumn{2}{c|}{Number of qubits} & \multicolumn{3}{c}{Number of quantum gates} \\ \hline
     & Support & Total & X & $\text{C}^2\text{X}$ & Total \\ \hline
    $Seq_1$ & $n-2$ & $2n-1$
    & 0 & $4n-8$ & $4n-8$ \\ \hline
    $Seq_2$ & $1 \leq n_{s} \leq (n-2)$ & $n+n_{s}+1$
    & Eq.(\ref{eq:cost_X_wave}) & Eq.(\ref{eq:cost_T_wave}) & Eq.(\ref{eq:cost_wave}) \\ \hline
    $Seq_3$ & $1$ & $n+2$
    & Eq.(\ref{eq:cost_X_ladder}) & Eq.(\ref{eq:cost_T_ladder}) &  Eq.(\ref{eq:cost_ladder}) \\ \hline
    \end{tabular}
    \label{tab:cost_stesso_pp}
\end{table}
According to the different main composition ideas of the second and third PP-Stessos, the number of costed X gates for the PP-Stesso based on the $Seq_2$ in Eq.~(\ref{eq:cost_X_wave}) is less than the number of X gates for PP-Stesso based on the $Seq_3$ in Eq.~(\ref{eq:cost_X_ladder}).
\begin{equation}
    \label{eq:cost_X_wave}
    \begin{aligned}
        \text{No. }\text{X} &=
        2n - 2n_{s} - 2 -2j, 1 \leq  n_s \leq n - 2,
        %\\ & \text{where }
        j = \lceil \frac{n - n_s - 1}{2} \rceil - \lfloor \frac{n - n_s - 1}{2} \rfloor
    \end{aligned}
\end{equation}
\begin{equation}
    \label{eq:cost_X_ladder}
    \text{No. }\text{X} =
    2k^2 - 2k + 4m - 4, n = \frac{k(k+1)}{2} + m, 0 \leq m \leq k, 2 \leq k
\end{equation}

In contrast, the number of standard Toffoli gates $\text{C}^2\text{X}$ for the PP-Stesso based on the $Seq_2$ in Eq.~(\ref{eq:cost_T_wave}) is more than the number of $\text{C}^2\text{X}$ gates for PP-Stesso based on the $Seq_2$ in Eq.~(\ref{eq:cost_X_ladder}).
\begin{equation}
    \label{eq:cost_T_wave}
    \begin{aligned}
        \text{No. C}^2\text{X} &=
    \left\{
    \begin{array}{ll}
         6n - 2n_{s} - 16 + 2j, &1 \leq  n_s < n - 2 \\
         4n-8,& n_s = n-2
    \end{array}
    \right.
    \\& \text{where } j = \lceil \frac{n - n_s - 1}{2} \rceil - \lfloor \frac{n - n_s - 1}{2} \rfloor
    \end{aligned}
\end{equation}
\begin{equation}
    \label{eq:cost_T_ladder}
    \text{No. C}^2\text{X} = 
    \left\{
    \begin{array}{cc}
       2k^2 + 4m - 4,   & 
       n = \frac{k(k+1)}{2} + m, 0 \leq m \leq k, k = 2
       \\
       2k^2+6k+ 4m- 18,  &
       n = \frac{k(k+1)}{2} + m, 0 \leq m \leq k, 3 \leq k
    \end{array}
    \right.
\end{equation}

However, the circuit sizes of these two structures based on $Seq_2$ and $Seq_3$ are the same when there is only one support qubit of the second PP-Stesso, as calculated in Eq.~(\ref{eq:cost_wave}) and in Eq.~(\ref{eq:cost_ladder}) respectively.
\begin{equation}
    \label{eq:cost_wave}
    \text{S} = \text{Eq.~(\ref{eq:cost_X_wave})} + \text{ Eq.~(\ref{eq:cost_T_wave})} = \left\{ \begin{array}{cc}
         8n - 4n_{s} - 18, & n_s \neq n -2  \\
        4n-8, & n_s = n-2 
    \end{array}
    \right.
\end{equation}
\begin{equation}
    \label{eq:cost_ladder}
    \begin{aligned}
        \text{S} &= \text{ Eq.~(\ref{eq:cost_X_ladder})} + \text{ Eq.~(\ref{eq:cost_T_ladder})} \\&=
    \left\{
    \begin{array}{cc}
       4k^2 -2k + 8m - 8,   & 
       n = \frac{k(k+1)}{2} + m, 0 \leq m \leq k, k = 2
       \\
       4k^2+4k+ 8m- 22,  &
       n = \frac{k(k+1)}{2} + m, 0 \leq m \leq k, 3 \leq k
    \end{array}
    \right.
    \end{aligned}
\end{equation}

Thus, when there is only one support qubit, the third PP-Stesso is more efficient than the second PP-Stesso.
If the limitation of the total number of qubits for a quantum circuit, the second PP-Stesso has the advantage of adjustable support qubits. Notice that when there is $(n - 2)$ support qubits, the second PP-Stesso is the same as the first PP-Stesso. 

% Here, for the clear comparison of these three different PP-Stessos, we also list the number of qubits and quantum gates in Table~\ref{tab:cost-stesso-pp-exp}.
% \begin{table}[h!]
%     \centering
%     \caption{Quantum cost of different PP-Stessos for an $(14+1)$-bit Toffoli gate (14 controls, one target) based on the combination sequences $Seq_1, Seq_2, Seq_3$ of the composed shaped structures.}
%     \begin{tabular}{p{2cm}|cc|ccc}
%     \hline
%     {Combination sequences} & \multicolumn{2}{c|}{Number of qubits} & \multicolumn{3}{c}{Number of quantum gates} \\ \hline
%      & Support & Total & X & $\text{C}^2\text{X}$ & Total \\ \hline
%     $Seq_1$ & $12$ & $27$
%     & 0 & $48$ & $48$ \\ \hline
%     $Seq_2$ & $12$ & $27$
%     & 0 & 48 & 48 \\ \hline
    
%     %  wave-shape when (13 - $n_s$) is even  & 
%     % $14$ & $[2, 11]$ & 1 & $n_{s}+15$
%     % & $26-2n_{s}$ & $48-2n_{s}$ & $74 - 4n_{s}$ \\ \hline
%     $Seq_2$ & $\{3,5,6,7,9,11\}$ & $n_{s}+15$
%     & $26-2n_{s}$ & $68-2n_{s}$ & $94 - 4n_{s}$ \\ \hline

%     %  wave-shape when (13 - $n_s$) is odd & 
%     % $14$ & $[2, 11]$ & 1 & $n_{s}+15$
%     % & $24 - 2n_{s}$ & $50 - 2n_{s}$ & $74 - 4n_{s}$ \\ \hline
%     $Seq_2$ & $\{2,4,6,8,10\}$ & $n_{s}+15$
%     & $24 - 2n_{s}$ & $70 - 2n_{s}$ & $94 - 4n_{s}$ \\ \hline
    
%     $Seq_2$ & $1$ & $16$
%     & 24 & 66 & $90$ \\ \hline 
%     $Seq_3$ & $1$ & $16$
%     & 36 & 54 & 90 \\ \hline
%     \end{tabular}
%     \label{tab:cost-stesso-pp-exp}
% \end{table}

As illustrated in Table~\ref{tab:cost_stesso_pp}, the complexity of the circuit sizes of three PP-Stesso based on $Seq_1, Seq_2, Seq_3$ is the same, i.e., always the polynomial complexity $O(n)$.
For the circuit depths of these three structures, the complexities of the circuit depths are also polynomial complexity $O(n)$, but the exact values are different. The circuit depth of the first PP-Stesso of $Seq_1$ is exactly its circuit size. For the second PP-Stesso of $Seq_2$, its circuit depth is less than its circuit size because of the circuit depths of the used one downstairs-shaped structure and one upstairs-shaped structure in Eq.~(\ref{eq:seq_wave}) are both $2\log(\lfloor \frac{n - n_s - 1}{2} \rfloor + 1)$. And the circuit depth of the third PP-Stesso of $Seq_3$ is greater than the second PP-Stesso because the circuit depths of all downstairs-shaped structures and upstairs-shaped structures in Eq.~(\ref{eq:seq_stair}) are both approximately $2\sqrt{2(n+1)}$.

Notice that the complexities of a V-shaped structure, a downstairs-shape structure that satisfies the rule of control $\rightarrow$ target in Eq.~(\ref{eq:seq_wave}), and a downstairs-shape structure that satisfies the rule of target $\rightarrow$ control in Eq.~(\ref{eq:seq_stair}) are $O(n)$, $O(log(n))$, and $O(\sqrt{n})$ respectively.

Since we have different decomposition methods of sub-product terms for an $(n+1)$-bit Toffoli gate with positive polarities, the MP-Stesso for an $(n+1)$-bit Toffoli gate with mixed polarities can be defined in Definition~\ref{def:Stesso-MP}, by adding or removing X gates into a PP-Stesso.
\begin{definition}
    \label{def:Stesso-MP}
    The step-decreasing
structures shaped operators are a conceptual design for representing an $(n+1)$-bit Toffoli gate with mixed polarities of all $n$ control qubits, such a design is named ``MP-Stesso". The MP-Stesso is a composed circuit based on a PP-Stesso by adding or removing X gates iff its related control qubits have negative polarities
   .
\end{definition}
\begin{property}
    \label{pro:stesso_mp_ds}
    The circuit depth and size of an MP-Stesso are the sum of its corresponding PP-Stesso and the number $n_{neg}$ of negation polarities of the sub-product term of control qubits.
\end{property}
According to the circuit depth and circuit size of an MP-Stesso in Property~\ref{pro:stesso_mp_ds}, the complexity of circuit depth and circuit size remains the same as its corresponding PP-Stesso. So we won't go into details here. If one wants to directly obtain a specific output of an $(n+1)$-bit Toffoli gate with mixed polarities of its sub-product terms, follow the two different composition ideas of the combination sequence $Seq_2$ and $Seq_3$, then the outputs will be in the similar form $\prod_{i=1}^{n_s}{q_{c_i}}\prod_{2j\in [n_s, n]}{q_{c_j}q_{c_{j+1}}}$ and $\prod_{i=1}^{2}{q_{c_i}}\prod_{i=3}^{2+3}{q_{c_i}\cdots\prod_{i = \frac{k(k+1)}{2}}^{\frac{k(k+1)}{2} + m}{q_{c_i}}}$, where the $\prod{q_{c_i}}$ can have mixed polarities such as $\overline{\overline{q_{c_1}q_{c_2}}q_{c_3}}q_{c_4}$ and ${q_{c_j}q_{c_{j+1}}}$ can have different polarities including $q_{c_j}q_{c_{j+1}}$ and $\overline{q_{c_j}q_{c_{j+1}}}$.

To extend the reconfigurability of Stesso to not only be constrained to polarities, there is a more general conceptual design in Definition~\ref{def:Stesso-G} based on MP-Stesso by using some constrained unitary gates.

\begin{definition}
    \label{def:Stesso-G}
    The step-decreasing
structures shaped operators are a conceptual design for representing a generalized $(n+1)$-bit Toffoli gate with output form of mixed OR and ESOP, such a design is named ``G-Stesso". The G-Stesso is a composed circuit based on an MP-Stesso by adding specific constrained unitary gates $U_i$. The final states of the composition circuit form top-to-bottom are
% \[
% U_2^{-1}(q_{c_j},q_{s_i},q_t)M'(q_{c_j},q_{s_i},q_t)U_1^{-1}(q_{c_1}, q_{c_2}, q_{s_1})M(q_{c_i},q_{s_i},q_t)U_2(q_{c_j},q_{s_i},q_t)U_1(q_{c_1}, q_{c_2}, q_{s_1})(q_{c_i},q_{s_i},q_t) = (q_{c_i},q_{s_i},q'_t)
% \]
 $U_2^{-1}(q_{c_j},q_{s_i},q_t)M'(q_{c_j},q_{s_i},q_t)U_1^{-1}$ $(q_{c_1}, q_{c_2}, q_{s_i})$ $M(q_{c_i},q_{s_i},q_t)U_2(q_{c_j},q_{s_i},q_t)$ $U_1(q_{c_1}, q_{c_2}, q_{s_i})(q_{c_i},q_{s_i},q_t) = (q_{c_i},q_{s_i},q'_t)$, where $3 \leq j \leq n$, $M$ and $M'$ are the two matrices obtain by the step-decreasing composed shaped structure of MP-Stesso.
 The constrained unitary gates $U_1$ satisfy all the following rules:
    \begin{enumerate}
        \item Only can relate to the first two controls and all support qubits $q_{c_1}, q_{c_2}$, $q_{s_i}$.
        \item Consists of a set of several X gates, CX gates, and $\text{C}^2\text{X}$.
    \end{enumerate}
   .
\end{definition}

As stated in Definition~\ref{def:Stesso-G}, this design for a more flexible output form is derived from the two unentangled unitary gates $U_1$ and $U_2$ and their reversed gates.
By inserting $U_1^{-1}$ into two step-decreasing composed shaped structures of an MP-Stesso, the controls of the second composed shaped structure remain the same as the first composed shaped structure. And for the states of all support qubits of the second composed structure become different from the previous structure. This also means that the first two controls will not affect the second composed shaped structure. A specific example of using a G-Stesso is the design of an four-bit binary comparator in section~\ref{sec:bc}.

\section{Application in a binary comparator}

% \subsection{Adder}

% \subsection{Subtractor}

% \subsection{binary comparator}
\label{sec:bc}
The $n$-bit binary comparator~\cite{mano2023digital} is used to fully compare two $n$-bit inputs $x_n \cdots x_1$ and $y_n \cdots y_1$, and produce three outputs 
%given in Eq.~(\ref{eq:comparator}).
if $x_n \cdots x_1 < y_n \cdots y_1$, $x_n \cdots x_1 = y_n \cdots y_1$ or $x_n \cdots x_1 > y_n \cdots y_1$.

Here, the truth table of a 1-bit binary magnitude comparator is provided as shown in Table~\ref{tab:comparator_1} for a better understanding of the relationships between the three outputs.

\begin{table}[h!]
    \centering
    \caption{The truth table of a 1-bit binary magnitude comparator.}
    \begin{tabular}{cc|ccc}
    \hline
     \multicolumn{2}{c|}{Inputs} & \multicolumn{3}{c}{Outputs $f$} \\ \hline
      $x_1$ & $y_1$ & $x_1 < y_1$ & $x_1 = y_1$ & $x_1 > y_1$ \\ \hline
    0 & 0 & 0 & 1& 0\\ \hline
    0 & 1 & 1 & 0& 0\\ \hline
    1 & 1 & 0& 1& 0\\ \hline
    1 & 0 & 0& 0&1 \\ \hline
    \end{tabular}
    \label{tab:comparator_1}
\end{table}
Obviously, the three outputs are $f(x_1 < y_1) = \overline{x}_1 y_1$, $f(x_1 = y_1) = \overline{x}_1 \oplus y_1$, and $f(x_1 > y_1) = x_1 \overline{y}_1$. Notice that $f(x_1 > y_1) = x_1 \overline{y}_1 = \overline{x}_1 y_1 \oplus x_1 \oplus y_1 = f(x_1 < y_1) \oplus \overline{f(x_1 = y_1)}$.

% Generally, the output function $f$ for an $n$-bit binary magnitude comparator can be obtained by Eq.~(\ref{eq:comparator}).
% \begin{equation}
%     \label{eq:comparator}
%     \begin{aligned}
%         f &= \left\{
%         \begin{array}{ll}
%             \bigoplus_{i=1}^{n}{\overline{x}_{n+1-i}y_{n+1-i} \prod_{j=2}^{i}{(\overline{x}_{n+2-j} \oplus y_{n+2-j})}}, & 
%             x_n \cdots x_1 < y_n \cdots y_1
%             \\
%             \prod_{i=1}^{n}{(\overline{x}_i \oplus y_i)}, & x_n \cdots x_1 = y_n \cdots y_1
%             \\
%             \bigoplus_{i=1}^{n}{{x}_{n+1-i}\overline{y}_{n+1-i} \prod_{j=2}^{i}{({x}_{n+2-j} \oplus \overline{y}_{n+2-j})}},
%             &
%             x_n \cdots x_1 > y_n \cdots y_1
%         \end{array}
%         \right.
%     \end{aligned}
% \end{equation}

In this paper, notations $f(x_n \cdots x_1 < y_n \cdots y_1)$, $f(x_n \cdots x_1 = y_n \cdots y_1)$, and $f(x_n \cdots x_1 > y_n \cdots y_1)$ represent the three outputs whether the input $x_n \cdots x_1$ is less than, equal to and greater than $y_n \cdots y_1$ respectively. The output $f(x_n \cdots x_1 > y_n \cdots y_1)$ equals the negation of the xor result of two outputs $f(x_n \cdots x_1 < y_n \cdots y_1)$ and output $f(x_n \cdots x_1 = y_n \cdots y_1)$, i.e., output $f(x_n \cdots x_1 > y_n \cdots y_1) = f(x_n \cdots x_1 $ < $ y_n \cdots y_1) \oplus \overline{f(x_n \cdots x_1 = y_n \cdots y_1)}$.

Take a 4-bit binary magnitude comparator as an example, the three outputs are given in Eq.~(\ref{eq:comparator_4}).
\begin{equation}
    \label{eq:comparator_4}
    \begin{aligned}
    f(x_4 x_3 x_2 x_1 < y_4 y_3 y_2 y_1) =&
    \bar{x}_4y_4 \oplus (\bar{x}_4 \oplus y_4)\bar{x}_3y_3 \oplus 
    \\&
    (\bar{x}_4 \oplus y_4)(\bar{x}_3 \oplus y_3)\bar{x}_2y_2
    \oplus \\&
    (\bar{x}_4 \oplus y_4)(\bar{x}_3 \oplus y_3)(\bar{x}_2 \oplus y_2)\bar{x}_1y_1
    \\
    \hspace{1.5cm}
    f(x_4 x_3 x_2 x_1 = y_4 y_3 y_2 y_1) =& (\bar{x}_4 \oplus y_4)(\bar{x}_3 \oplus y_3)(\bar{x}_2 \oplus y_2)
    (\bar{x}_1 \oplus y_1)
    \\
    \hspace{1.5cm}
    f(x_4 x_3 x_2 x_1 > y_4 y_3 y_2 y_1) =&
    x_4\bar{y}_4 \oplus (x_4 \oplus \bar{y}_4)x_3\bar{y}_3 \oplus 
    \\&
    (x_4 \oplus \bar{y}_4)(x_3 \oplus \bar{y}_3)x_2\bar{y}_2
    \oplus \\&
    (x_4 \oplus \bar{y}_4)(x_3 \oplus \bar{y}_3)(x_2 \oplus \bar{y}_2)x_1\bar{y}_1
    \end{aligned}
\end{equation}

The two outputs $f(x_4 x_3 x_2 x_1 < y_4 y_3 y_2 y_1)$ and $f(x_4 x_3 x_2 x_1 > y_4 y_3 y_2 y_1)$ are in similar form, with the only difference of the polarities of $x_i$ and $y_i$, where $1 \leq i \leq 4$. Notice that there are several different lengths of product terms of inputs $x_i$ and $y_i$, which can be obtained by using $(n+1)$-bit Toffoli gates. The maximum length of all product terms in output $f(x_4 x_3 x_2 x_1 < y_4 y_3 y_2 y_1)$ is five, i.e., $(\bar{x}_4 \oplus y_4)(\bar{x}_3 \oplus y_3)(\bar{x}_2 \oplus y_2)\bar{x}_1y_1$. 
And the output $f(x_4 x_3 x_2 x_1 < y_4 y_3 y_2 y_1)$ can be transformed into a similar form of the output of a V-shape structure (consisting of five controls $q_{c_i}$, three supports $q_{s_i}$, and one target $q_t$) in Definition~\ref{def:v-shape}, as given in Eq.~(\ref{eq:comparator_4_le}).
\begin{equation}
\label{eq:comparator_4_le}
    \begin{aligned}
    &f(x_4 x_3 x_2 x_1 < y_4 y_3 y_2 y_1) \\=& \;
    % \bar{x}_4y_4 \oplus (\bar{x}_4 \oplus y_4)\bar{x}_3y_3 \oplus 
    % %\\&
    % (\bar{x}_4 \oplus y_4)(\bar{x}_3 \oplus y_3)\bar{x}_2y_2
    % \oplus 
    % \\&
    % (\bar{x}_4 \oplus y_4)(\bar{x}_3 \oplus y_3)(\bar{x}_2 \oplus y_2)\bar{x}_1y_1
    % \\=& \;
    % (\bar{x}_4 \oplus y_4)((\bar{x}_3 \oplus y_3)((\bar{x}_2 \oplus y_2)(\bar{x}_1 y_1 \oplus y_2) 
    % \oplus y_2 \oplus y_3) \oplus y_3 \oplus y_4) \oplus y_4
    % \\=& \;
    q_{c_5}(q_{c_4}(q_{c_3}(q_{c_2}q_{c_1} \oplus q_{s_1}) \oplus q_{s_2}) \oplus q_{s_3}) \oplus q_t,
    \\
    &\; %\text{where }
    q_{c_5} = \bar{x}_4 \oplus y_4, \;
    q_{c_4} = \bar{x}_3 \oplus y_3, \;
    q_{c_3} = \bar{x}_2 \oplus y_2, \;
    q_{c_2} = \bar{x}_1,
    q_{c_1} = y_1,
    \\& \; %\qquad \quad
    q_{s_3} = y_3 \oplus y_4, \;
    q_{s_2} = y_2 \oplus y_3, \;
    q_{s_1} = y_2, \;
    q_t = y_4
    \end{aligned}
\end{equation}
% \\& \qquad \qquad \qquad \qquad \;

% \textbf{CNF(conjunction normal form)-XOR SAT logical structure}

\begin{figure}[h!]
    \centering
    \includegraphics[width=0.9\linewidth]{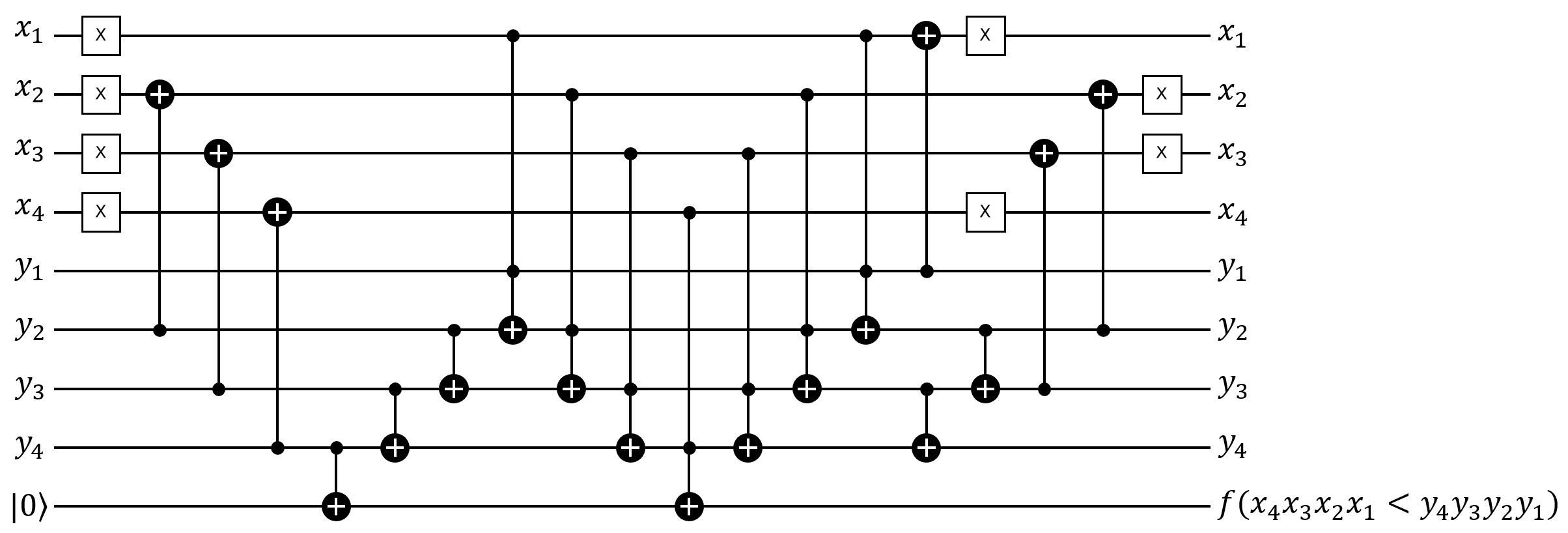}
    \caption{Quantum circuit of the output $f(x_4 x_3 x_2 x_1 < y_4 y_3 y_2 y_1)$ of a 4-bit binary magnitude comparator.}
    \label{fig:BC_4bit_lt}
\end{figure}

\begin{figure}[h!]
    \centering
    \includegraphics[width=0.95\linewidth]{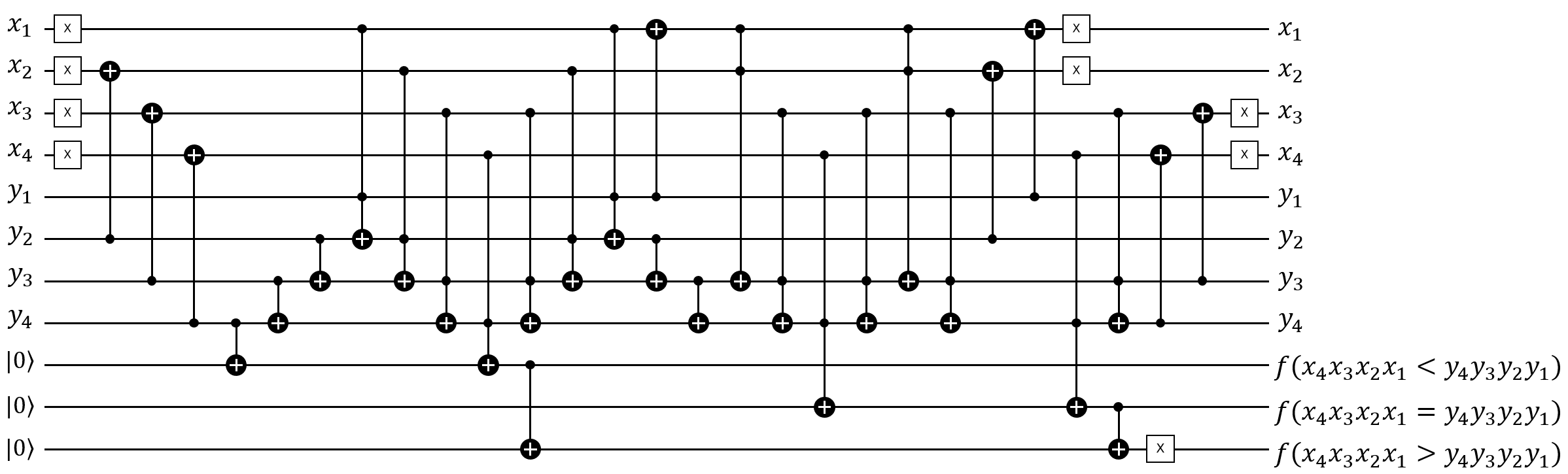}
    \caption{Final quantum circuit of a 4-bit binary magnitude comparator.}
    \label{fig:BC_4bit}
\end{figure}

Since all states of all controls, supports, and one target in Eq.~(\ref{eq:comparator_4_le}) can be implemented utilizing X gates and CX gates, the quantum circuit of one output $f(x_4 x_3 x_2 x_1 < y_4 y_3 y_2 y_1)$ of a 4-bit binary magnitude comparator is stated in Figure~\ref{fig:BC_4bit_lt}. 
Based on Eq.~(\ref{eq:comparator_4}), the output $f(x_4 x_3 x_2 x_1 = y_4 y_3 y_2 y_1)$ can be directly obtain using the quantum circuit of an $(4+1)$-bit Toffoli gate. According to the relationship among all three outputs of a 4-bit binary comparator, the final quantum circuit is given in Figure~\ref{fig:BC_4bit}.

% Notice that there is a reversed part of the contained quantum gates between the fourth stage and fifth stage, the quantum circuit can be further simplified as shown in Figure~\ref{fig:BC_4bit}. 

% \begin{figure}[h!]
%     \centering
%     \includegraphics[width=0.95\linewidth]{figs/BC_4bit_org.png}
%     \caption{Quantum circuit of a 4-bit binary magnitude comparator.}
%     \label{fig:BC_4bit_org}
% \end{figure}

% Generally, an $n$-bit binary comparator can be designed by following the same methodology. The output function $f(x_n \cdots x_1 < y_n \cdots y_1)$ can be obtained by Eq. ().

% \subsection{Maslov cost for theoretical decomposition}

% \textbf{use IBM's decomposition}

% \textbf{use GALA and CALA}

\section{Conclusions}

This paper introduced three types of step-decreasing structures shaped operators ``Stesso", with positive polarities ``PP-Stesso", mixed polarities ``MP-Stesso", and generalized ``G-Stesso" of mixed SOP and ESOP logical structure, for representing $(n+1)$-bit Toffoli gates, where $n \geq 3$.
 % Notice that SOP is a fundamental structural form for classical design, while ESOP is a fundamental structural form for quantum design.
 % (PP-Stesso, MP-Stesso, and G-Stesso)  
 
Three types of Stesso are conceptual designs by composing different basic logically shaped structures, including symbol shapes with their extensions (such as \drawdownstairs{1}-shape), English shapes (such as V-shape), and Cyrillic shapes (such as {\foreignlanguage{russian}{И}}-shape). By using the standard 3-bit Toffoli gates $\text{C}^2\text{X}$, X gates, and some constrained unitary gates like CX gates, our Stesso circuits are reconfigurable and layout-aware with adjustable support qubits.

Specifically, the Stesso can be directly used to generate an output in specific mixed SOP and ESOP form, such as $q_{c_5}(q_{c_4}(q_{c_3}(q_{c_2}q_{c_1} \oplus q_{s_1}) \oplus q_{s_2}) \oplus q_{s_3}) \oplus q_t$ and $\overline{\overline{q_{c_1}q_{c_2}}q_{c_3}q_{c_4}}$ $\overline{q_{c_5}q_{c_6}}q_{c_7} \oplus q_t = \overline{(\overline{q_{c_1}} + \overline{q_{c_2}})q_{c_3}q_{c_4}}\overline{q_{c_5}q_{c_6}}q_{c_7} \oplus q_t$. The circuit size for implementing our Stesso has the polynomial complexity of $O(n)$. The complexity of its circuit depth has a maximum $O(n)$ and an approximate minimum $O(log(n))$. Moreover, the composed-shape structure of a Stesso circuit is always symmetrical to obtain a single output, which makes this Stesso well-suited for symmetrical-based structures of real QPUs. Additionally, we used Stesso to directly design an $n$-bit binary comparator, without using ancilla and combination units of each 2-bit binary comparator of conventional design methods. Concluding, our design methodology proves that Stesso can be utilized in the practical construction of arithmetic units (such as in arithmetic applications~\cite{li2020efficient} and quantum search algorithms~\cite{hou2020quantum,gao2021novel}), by maintaining an effective structure design of fewer involved qubits, circuit sizes, and circuit depths.

% \textbf{Mention the methodology of inserting Pauli-X gates into a quantum circuit. (Demorgan)}
% \textbf{Mixed of SOP and ESOP for output}

% \textbf{Questions: optimization methods of mixed SOP and ESOP?}

% \textbf{$(n+1)$-bit Toffoli gates is a generalization of ESOP.}

% Mention our three proposed construction advantages, which are customization, layout-awareness, and using support qubits.

% \textbf{Multiplier}

\bibliographystyle{ieeetr}
\bibliography{ref}

\begin{thebibliography}{10}

\bibitem{bub2010quantum}
J.~Bub, ``Quantum computation: Where does the speed-up come from,'' {\em Philosophy of quantum information and entanglement}, pp.~231--246, 2010.

\bibitem{childs2018toward}
A.~M. Childs, D.~Maslov, Y.~Nam, N.~J. Ross, and Y.~Su, ``Toward the first quantum simulation with quantum speedup,'' {\em Proceedings of the National Academy of Sciences}, vol.~115, no.~38, pp.~9456--9461, 2018.

\bibitem{larose2019overview}
R.~LaRose, ``Overview and comparison of gate level quantum software platforms,'' {\em Quantum}, vol.~3, p.~130, 2019.

\bibitem{tan2020optimal}
B.~Tan and J.~Cong, ``Optimal layout synthesis for quantum computing,'' in {\em Proceedings of the 39th International Conference on Computer-Aided Design}, pp.~1--9, 2020.

\bibitem{chamberland2020topological}
C.~Chamberland, G.~Zhu, T.~J. Yoder, J.~B. Hertzberg, and A.~W. Cross, ``Topological and subsystem codes on low-degree graphs with flag qubits,'' {\em Physical Review X}, vol.~10, no.~1, p.~011022, 2020.

\bibitem{ibm2025computer}
IBM, ``Ibm quantum platform: Compute resources.'' \url{https://quantum.cloud.ibm.com/computers}, 2025.
\newblock Accessed: 2025-09-22.

\bibitem{toffoli1980reversible}
T.~Toffoli, ``Reversible computing,'' in {\em International colloquium on automata, languages, and programming}, pp.~632--644, Springer, 1980.

\bibitem{barenco1995elementary}
A.~Barenco, C.~H. Bennett, R.~Cleve, D.~P. DiVincenzo, N.~Margolus, P.~Shor, T.~Sleator, J.~A. Smolin, and H.~Weinfurter, ``Elementary gates for quantum computation,'' {\em Physical review A}, vol.~52, no.~5, p.~3457, 1995.

\bibitem{maslov2003improved}
D.~Maslov and G.~W. Dueck, ``Improved quantum cost for n-bit toffoli gates,'' {\em Electronics Letters}, vol.~39, no.~25, pp.~1790--1791, 2003.

\bibitem{orts2022studying}
F.~Orts, G.~Ortega, and E.~M. Garz{\'o}n, ``Studying the cost of n-qubit toffoli gates,'' in {\em International Conference on Computational Science}, pp.~122--128, Springer, 2022.

\bibitem{claudon2024polylogarithmic}
B.~Claudon, J.~Zylberman, C.~Feniou, F.~Debbasch, A.~Peruzzo, and J.-P. Piquemal, ``Polylogarithmic-depth controlled-not gates without ancilla qubits,'' {\em Nature Communications}, vol.~15, no.~1, p.~5886, 2024.

\bibitem{QiskitLattice}
Q.~Ecosystem, ``Lattice models - qiskit nature 0.7.2.'' \url{https://qiskit-community.github.io/qiskit-nature/tutorials/10_lattice_models.html}, 2025.
\newblock Accessed: 2025-09-26.

\bibitem{yu2024symmetry}
D.~Yu and K.~Fang, ``Symmetry-based quantum circuit mapping,'' {\em Physical Review Applied}, vol.~22, no.~2, p.~024029, 2024.

\bibitem{garvin2025data}
D.~Garvin, M.~Fiorentini, O.~Kondratyev, and M.~Paini, ``Data anonymisation with the density matrix classifier,'' {\em Cryptology ePrint Archive}, 2025.

\bibitem{mano2023digital}
M.~Morris and M.~D. Ciletti, {\em Digital design (6 ed.)}.
\newblock Pearson Educaci{\'o}n, 2023.

\bibitem{tyagi2020high}
P.~Tyagi and R.~Pandey, ``High-speed and area-efficient scalable n-bit digital comparator,'' {\em IET Circuits, Devices \& Systems}, vol.~14, no.~4, pp.~450--458, 2020.

\bibitem{al2009closed}
A.~N. Al-Rabadi, ``Closed-system quantum logic network implementation of the viterbi algorithm,'' {\em Facta universitatis-series: Electronics and Energetics}, vol.~22, no.~1, pp.~1--33, 2009.

\bibitem{thapliyal2010design}
H.~Thapliyal, N.~Ranganathan, and R.~Ferreira, ``Design of a comparator tree based on reversible logic,'' in {\em 10th IEEE International Conference on Nanotechnology}, pp.~1113--1116, IEEE, 2010.

\bibitem{donaire2024lowering}
L.~M. Donaire, G.~Ortega, E.~M. Garz{\'o}n, and F.~Orts, ``Lowering the cost of quantum comparator circuits,'' {\em The Journal of Supercomputing}, vol.~80, no.~10, pp.~13900--13917, 2024.

\bibitem{hausdorff2021set}
F.~Hausdorff, {\em Set theory}, vol.~119.
\newblock American Mathematical Soc., 2021.

\bibitem{rigetti2025QCS}
R.~Computing, ``Rigetti systems: Cepheus-1-36q quantum processor and ankaa-3 quantum processor.'' \url{https://qcs.rigetti.com/qpus}, 2025.
\newblock Accessed: 2025-09-26.

\bibitem{arute2019quantum}
F.~Arute, K.~Arya, R.~Babbush, D.~Bacon, J.~C. Bardin, R.~Barends, R.~Biswas, S.~Boixo, F.~G. Brandao, D.~A. Buell, {\em et~al.}, ``Quantum supremacy using a programmable superconducting processor,'' {\em Nature}, vol.~574, no.~7779, pp.~505--510, 2019.

\bibitem{google2025quantum}
G.~Q. AI and Collaborators, ``Quantum error correction below the surface code threshold,'' {\em Nature}, vol.~638, no.~8052, pp.~920--926, 2025.

\bibitem{chang2021mapping}
K.-Y. Chang and C.-Y. Lee, ``Mapping nearest neighbor compliant quantum circuits onto a 2-d hexagonal architecture,'' {\em IEEE Transactions on Computer-Aided Design of Integrated Circuits and Systems}, vol.~41, no.~10, pp.~3373--3386, 2021.

\bibitem{hetenyi2024creating}
B.~Het{\'e}nyi and J.~R. Wootton, ``Creating entangled logical qubits in the heavy-hex lattice with topological codes,'' {\em PRX Quantum}, vol.~5, no.~4, p.~040334, 2024.

\bibitem{li2020efficient}
H.-S. Li, P.~Fan, H.~Xia, H.~Peng, and G.-L. Long, ``Efficient quantum arithmetic operation circuits for quantum image processing,'' {\em Science China Physics, Mechanics \& Astronomy}, vol.~63, no.~8, p.~280311, 2020.

\bibitem{hou2020quantum}
W.~Hou and M.~A. Perkowski, ``Quantum-based algorithm and circuit design for bounded knapsack optimization problem.,'' {\em Quantum Inf. Comput.}, vol.~20, no.~9\&10, pp.~766--786, 2020.

\bibitem{gao2021novel}
P.~Gao, M.~A. Perkowski, Y.~Li, and X.~Song, ``Novel quantum algorithms to minimize switching functions based on graph partitions,'' {\em Cmc-Computers Materials \& Continua}, vol.~70, no.~3, p.~4545, 2021.

\end{thebibliography}

\end{document}